\documentclass[12pt,preprint,floatfix,aps]{revtex4}
\usepackage{graphicx,float}

\begin{document}
{\Large \bf \flushleft Risk trading, network topology, and banking regulation}

\vspace{0.5cm}

{\large \flushleft 
Stefan Thurner$^{1}$, Rudolf Hanel$^{2}$, and Stefan Pichler$^{3}$ }
{\small \flushleft 
  $^{1}${\it Institute of Mathematics, NuHAG, and HNO, Universit\"at Wien; Austria} \\
  $^{2}${\it Institute of Biomedical Engineering and Physics, 
          Universit\"at Wien; Austria}  \\
  $^{3}${\it Department of Finance, Technische Universit\"at Wien, Wien; Austria} \\}  
\vspace{0.8cm}

\begin{center}
{\small \bf Abstract}
\end{center}

\noindent
{\small In the context of understanding the nature of the risk 
transformation process of the financial
system we propose an iterative risk-trading game between several
agents who build their trading strategies based on a general 
utility setting. The game is studied numerically for different 
network topologies. Consequences of topology are shown for
the wealth time-series of agents, for the safety and efficiency 
of various types of networks. 
The proposed setup allows an analysis of the effects of different 
approaches to banking regulation as currently suggested by the Basle 
Committee of Banking Supervision. We find a phase transition-like 
phenomenon, where the Basle parameter 
plays the role of temperature and system safety serves as the 
order parameter. This result suggests the 
existence of an optimal regulation parameter. As a consequence 
a tightening of the current regulatory framework does not 
necessarily lead  to an 
improvement of the safety of the banking system. 
Moreover, we show that 
banking systems with local risk-sharing cooperations 
have higher global default rates than systems with low 
cyclicality. 

\vspace{1cm}
}

\vfill
\noindent
{\footnotesize {\bf Correspondence to:}\\
 Stefan Thurner;  HNO, AKH-Wien, University of Vienna\\
 W\"ahringer G\"urtel 18-20; A-1090 Vienna, Austria\\
 Tel.: ++43 1 40400 2099;  Fax: ++43 1 40400 3332\\
 e-mail: thurner@univie.ac.at
}  

\newpage 

\section{Introduction}

The efficiency and stability of the financial system and 
its institutions is seen as one of the core elements of  
modern economy. The regulation of financial intermediaries 
is thus a central issue which raises a number of  
questions both for practical implementations and for 
academic research. During the past two decades an intensive 
discussion about the regulation of financial markets 
evolved mostly driven by media attended events like the 
''Russian crisis'' in 1998 and defaults of big financial 
institutions like the 
failure of Barings Bank or, more recently, Long Term 
Capital Management. This discussion formed the basis 
for a regulatory innovation process under the lead of 
the Basle Committee of Banking Supervision which sets 
the relevant legal regulatory framework for the G-10 
economies as well as for the EU. As a consequence of 
the US Savings \& Loans disaster in the mid-eighties 
the first milestone of this process was the first Capital 
Accord in 1988 (Basle I) which forced banks to hold a capital cushion
of  8 \% of total assets to prevent default due 
to extreme unexpected losses. This was the first step 
where the activities of financial intermediaries were 
limited by the risk (measured by total assets) of their 
positions. Since total assets is an insufficient measure 
of risk at least for trading positions in securities 
and derivatives, the 1995 Amendment to the Capital 
Accord introduced a more risk-sensitive framework for 
market risk activities. Finally, the second Capital 
Accord (Basle II) which currently is under consultation 
is aimed to introduce a risk-sensitive framework for 
credit risk as well.

Although the discussion about the cornerstones of the 
new regulatory framework is in its final phase, there 
remains a number of open questions. 
Firstly, if the activities of financial intermediaries 
are limited by the risk of their current positions,
how should risk be measured? Secondly, if a capital 
cushion relative to the risk run by a bank is needed, 
what is its optimal size? Further, does a 
specific risk-sensitive framework lead to undesirable 
pro-cyclical effects in the economy? Finally, is this 
risk-regulatory framework sufficient to prevent the 
financial system from a total collapse when catastrophic 
events like September 11, 2001, occur?

It is the main purpose of this paper to address these 
questions and to develop a formal framework which allows 
a stringent analysis of the effects of different approaches 
of banking regulation. Moreover, the potential impact of 
catastrophic events on the financial 
system can be measured within this framework. To achieve 
this goal we propose an iterative risk-trading game between 
several agents who build their trading strategies based 
on a very general utility setting. 
The financial system is modeled using different network 
topologies which enables us to explicitly measure the 
network-topology dependence of different regulatory actions 
and various other important factors. This is a major 
contribution to the existing research because to our 
knowledge this is one of the first attempts of an analysis of 
financial markets where dynamic game models and network-topology 
are combined. Differing effects of single bank defaults and 
regional banking crises can easily be captured by our model. 
Additionally, the chosen network topology can be used to 
mimic specific structures of several national banking systems 
like ''sectors'' or risk-sharing cooperations . 

This research is related to the discussion about systemic 
risk in the banking sector, see 
\cite{Angelini96,Eisenberg01,Lehar01, Elsinger02}. 
The authors of \cite{Angelini96} empirically examine systemic 
risk in the Italian inter-bank clearing network where 
they explicitly address the potential size of a ''domino effect'' 
where the default of a single bank may jeopardize the 
ability of ''neighbor'' banks to meet their obligations. 
A first attempt to model systemic risk in financial networks 
was undertaken in \cite{Eisenberg01} where the authors show comparative 
statics describing the relationship between the clearing vector 
and the underlying parameters of the financial system which is 
exposed to external stochastic shocks. However, they model 
the behavior of financial institutions purely deterministically 
and they do not examine potential effects of network topology. 
A third stream of research, e.g. \cite{Lehar01}, focuses on the 
default correlation of banks as the central parameter of 
measuring systemic risk or contagion in the banking sector. 
The framework suggested in this paper enables us to analyze 
contagion as well as ''domino effects'' and provides a profound 
basis for comparative statics with respect to the architecture 
of the financial market combined with the regulatory framework. 
In a recent paper \cite{Elsinger02} an empirical application of
an extended version of \cite{Eisenberg01} is presented 
employing bank balance sheet data.
Their major finding is that contagious default from 
inter-bank relations plays only a minor role when risk is 
assessed at the level of the entire banking system.
This is an interesting contribution to the discussion 
about the scope and possible disadvantages 
of international banking regulation.

This paper is also related to research focused on the 
implications of the risk sensitive minimum
regulatory capital set by the  Basle II' Capital Accord. 
The role of  ''buffer capital'', i.e., free
capital held by banks that exceeds the minimum regulatory 
level, is addressed in \cite{Milne01}. In a dynamic model 
with endogenous capital they find that under reasonable
assumptions changes in capital requirements have no impact 
on bank behavior. Studies related to the possible pro-cyclical 
effect of risk-sensitive capital regulation
like \cite{Altman02,Catarineu02} find empirical evidence 
for pro-cyclicality at least for some widely used credit rating 
frameworks. From a purely theoretical perspective a macro-economic
model recently proposed in \cite{Baglioni02} demonstrates that 
in the short-run changes in capital regulation may have  
perverse effects (increasing supply for riskier borrowers when
restrictions are tightened and vice versa). 
In the long-run, however, these effects are absent and
bank capital regulation serves as a perfect tool for monetary 
policy transmission.

The remainder of the paper is organized as follows. 
Section 2 describes the model proposed to
address the central questions raised in the 
introductory section. The main results of the
numerical analysis are presented in section 3. 
Section 4 contains a brief discussion of the results
and concludes the paper.

\section{The Model}
\subsection{Model Structure} 
The aim of this work is to develop a 
simple model of a network of banks or financial 
institutions who can trade risk amongst each other.
Banks are forced to enter risk through demands of their clients, 
but try to compensate or share it amongst each other. 
We introduce an iterative dynamical model where agents (banks) can 
choose between different trading strategies, depending on their need to 
reduce individual risk. The need to reduce risk depends on the 
wealth of a bank and regulatory parameters.  
Risk trading actions take place between players whose connections 
are characterized by an exogenously specified graph  topology. This topology 
remains constant over time in the current setting, i.e., the
number of banks and the structure of the trading network can 
change endogenously only due to bank defaults.
We introduce measures which characterize the 
performance of different network topologies with 
respect to overall fairness and stability of the network,  
individual safety, i.e., survival probability, and the 
efficiency of reducing exogenous, i.e., 
externally enforced, risk.  
By studying the time evolution of the wealth of interacting 
players we find remarkable coincidence with realistic 
financial time-series. Characteristics of those time-series 
depend on topology and regulation parameters. 
It is important to mention here that in contrast to some recent 
approaches in cooperative network- and game theory, see e.g.  
\cite{Challet97,Suijs98,Savit99,Burt99},  
our agents do not learn but are purely selfish. In our present setting 
they do not have any incentive to adapt for cooperation.   

The specific model analyzed in this paper restricts the 
decision of the agents only to depend on
their individual risk rather than to their assessment 
of their counterpartie's risk. If agents take the
risk of their counterparties into account the terms of 
trade (e.g. bid-ask spreads) of the riskier
agents will deteriorate and as a consequence a 
self-reinforcing dynamics will increase the risks
of these agents and their default probabilities. 
Recent empirical evidence, like the default of
Enron, supports this reasoning.  The simplified 
framework of this paper, however, enables us to
analyze some realistic situations. Firstly, the risk 
assessment of counterparties does not change
continuously but in discrete time steps, e.g. when 
a rating change is announced. Thus, for
shorter periods of time (particularly during intra-day periods) 
changes in the risk of an agent's
position will not change its terms of trade immediately. 
Secondly, the regulatory framework currently in discussion 
is aimed to create a  ''level playing-field'' where agents can 
safely interact without permanent risk assessment of their 
counterparties. This is exactly related to our
simplified framework. We have to emphasize that the inclusion 
of risk assessment by agents is a valuable extension 
of the framework presented in this paper but is left for
future research.

In the following we deal with a set of $N$ sites (representing  
banks) labeled by $i=1,\cdots,N$ who receive requests from their 
$N$ exterior clients.  For simplicity each bank has only one client. 
Note, that no other interactions of the financial sector (banks) 
and the real sector 
of the economy (clients) are endogenously modeled in this framework.
The client's requests (forwards, swaps, weather derivatives, etc.) 
are modeled by bets which depend on an external random process $X(t)$, 
say the weather, and a time of maturity $t_m$. 
For the rest of this paper we consider a simple binary random process 
$X(t)\in\{-1,1\}$ with equal probabilities 
$p(-1) =p(1)=\frac{1}{2}$.   
Note,  that the model can be easily extended to cover non-iid-processes like 
cyclical processes by changing the specification of the dynamics of $X(t)$. 
 
For example, a client today bets 
20 m USD that on $t_m={\rm Dec.~15^{th}~2004}$ there will be rain. 
In this model, the bank is forced to accept these external bets 
and receives a spread or incentive, $\delta^{\rm exo}$, as a 
compensation. This incentive may be
interpreted as a fixed fee or a customer related bid-ask spread. 
At each time-step in the game, exterior clients confront their 
individual banks with external bets with 
uniformly distributed times of maturity,  
$t_m=t+\tau,~\tau\in\{0,1\cdots,  \tau^{\rm max}\}$, where 
$t$ denotes present time.  For two players only, 
compare e.g. with \cite{Schlesinger84,Gollier96}. 
The associated betting volume offered to bank $i$ is a 
random number also drawn from a uniform distribution 
$B^{exo}_i(t+\tau)\in[-D(t) \cdot W_i(t),D(t) \cdot W_i(t)]$, 
where $W_i(t)$ is the present wealth of bank $i$. 
This means that the maximum betting volume asked from a bank is 
proportional to its present wealth 
(the smaller the  banks the smaller their transactions).
$D$ is the fraction of external risk a bank is allowed to take 
within a given time-step, 
with respect to its own wealth. 
$D$ could be seen as a regulatory parameter since it imposes an
upper limit for a bank's risk dependent on its wealth 
(or equity in a wider sense). This parameter does not yet resemble the 
8 \%-rule currently set by regulators, where some risk measure (e.g.,
risk-weighted nominal value of positions) must not exceed the 
12.5 multiple of a bank's equity. Rather, $D$ can be understood 
as a measure that prevents banks to accept large risks within short 
time scales. The 8 \%-rule will be incorporated in the model in the 
next section. 
Since in case of no interactions, $D$  would simply be the 
volatility of wealth processes, 
in the following we shall call $D$ the exogenous volatility.  

By accepting a bet (which matures in $\tau$ days from now) 
from a client, bank $i$ enters risk 
$R_i^{exo}(t+\tau)$, which is defined as 
the maximum amount the bank can loose at time $t+\tau$, 
i.e., $R_i^{exo}(t+\tau)=|B^{exo}_i(t+\tau)|$.  
This risk measure is related to a great variety of risk measures 
used by regulators, e.g. the standard deviation of profit and 
losses (volatility), and the 1 \%-quantile of the profit and loss
distribution (value-at-risk).

\subsection{Model Dynamics}
The essence of the model is that this risk ($R_i^{exo}$) can be   
traded away, if the bank is able to find a neighbor bank, 
which is willing to enter a betting contract which serves to reduce 
this risk. In the above example, 
the bank will either look if an appropriate  bet is already offered 
on the ''market'' (issued by a neighbor bank), 
or it will issue one itself to its neighbor sites, which would read:   
20 m USD that on $t_m={\rm Dec.~15^{th}~2004}$ there will be sunshine. 
All the contracts are kept in the  ''betting book''  $B_{ij}(t+\tau)$, 
which contains the betting volume bank $i$ 
is betting against bank $j$ at time $t+\tau$. 
Note, that a bet which one bank wins the other loses, 
i.e., $B_{ij}(t+\tau)=-B_{ji}(t+\tau)$. 
If bank $i$ offers a bet maturing at $t+\tau$ it does so through an entry in the 
''market'' book $M_{i}(t+\tau)$. The numerical value of $M_i$ is the volume.  
The market book is ''visible'' to all neighbor banks of $i$ only. 
Whenever a bank agrees to a bet, the offering party is obliged to pay 
a spread, $\delta$, to the accepting bank.   
In reality $\delta$ is a dynamical 
quantity which is determined on the market. Here, for simplicity, 
we choose $\delta$ to be 10 \% of the bet volume $B_{ij}(t+\tau)$.
After entering a contract, risk of bank $i$ is defined as the maximum 
amount he can loose at time $t+\tau$, and now consists of the external bets 
and all the bets entered with neighbor sites, i.e.,  
\begin{equation}
R_i(t+\tau)= \left| B^{exo}_i(t+\tau) + \sum_{j} B_{ij}(t+\tau) \right| \quad , 
\end{equation}
where $j$ indexes the neighboring banks.   
To explicitely model the decision making process we assume that 
each bank $i$ is equipped with a von Neumann-Morgenstern utility 
function with risk aversion depending on a
parameter $\alpha_{i} > 0$. To model heterogeneity of banks we 
model the risk aversion parameter as a random variable in 
our framework. From a regulator's point of view
this makes sense because in reality there is a non-negligible 
uncertainty about the risk aversion
parameters of banks either across the cross-section or across time.

There are two distinct kinds of decisions banks can make. 
The first situation arises if a bank $i$ finds a bet (with maturity at $t+\tau$) 
on the market ($M_{j}(t+\tau)$ 
offered by a neighbor site $j$) which serves to reduce 
its risk, i.e., 
$\left|B^{exo}_i(t+\tau)+ \sum_j B_{ij}(t+\tau)+M_{j}(t+\tau) \right| 
<  R_i(t+\tau) $, it will obviously accept it and 
receive the spread for the deal. 
This we  call the {\it passive strategy}. On average, the bank
will gain the spread for sure and then await the outcome of 
the bet, where there is a fair 1:1 chance to win. 
In this situation a bank can increase its expected utility and
simultaneously reduce its risk. Thus, this decision does not 
depend on $\alpha_i$ and the specific choice of the utility 
function is irrelevant, as long as positive marginal utility and 
risk aversion are ensured (which both follow from the von 
Neumann-Morgenstern property).

In the second situation there is no (risk-reducing) 
offer in the market book for any given date of maturity $t+\tau$, 
and the bank has to decide to either do nothing or to offer a bet 
itself to its neighbor banks (for that date), which we call the 
{\it active strategy}. The decision now depends on the present 
wealth $W_i(t)$, the risk of the bank $R_i(t+\tau)$, the 
spread $\delta$ and the risk aversion parameter $\alpha$. 
Since agents have to decide whether to adopt the risk-reducing 
active strategy for each maturity $t+\tau$
separately, we are interested in the 
random outcome of the decision represented by 
the probability of adopting 
such a strategy which we denote with $p_i^{\rm active}(t,\tau)$. 

To simplify the decision making process in a numerical framework 
we avoid the explicit maximization of expected utility 
and directly model the outcome of the process by assuming
that
\begin{equation} 
 p_i^{\rm active}(t,\tau) =   
1-{\rm exp}\left[- \frac{\alpha_i \, R_i(t+\tau)}{W_i(t)+\delta} \right]  
\quad . 
\label{prob_act}
\end{equation} 
Note, that this structure is consistent with the general 
von Neumann-Morgenstern setting, since $p_i^{\rm active}(t,\tau)$
is increasing in $\alpha_i$ and $R_i(t+\tau)$ and decreasing in 
$W_i(t)$ and $\delta$, respectively. 

The interaction between banks and the decision making is thus 
not fully deterministic. 
Here we kept the risk aversion factor constant, $\alpha_i=5$, for all 
banks in all the following computations. 
In contrast to previous work, e.g. \cite{Eisenberg01}, 
we introduce a  noise component to the behavior of agents 
which replaces the role of the utility function in the decision 
making process. Whereas from a single-agent perspective this 
might be regarded as an arbitrary approximation, from a 
regulator's perspective, however, this construction is realistic, 
because the regulator faces a large number of agents with different 
utility functions which are not known by the regulator. 
Thus, our decision making rule (\ref{prob_act}) 
explicitly models the uncertainty about the single-agent 
utility functions, or put equivalently, 
models the stochastic deviations of the decisions of  
single agents from the representative agent.
Note, however, that the introduction of decision making rule 
(\ref{prob_act}) serves only as a simplification
of a purely utility based framework and does not lead 
to any loss of generality.

To model the Basle regulatory framework, the probability for 
adopting an active strategy is always  equal to one, whenever 
the total risk exceeds a certain percentage of wealth, i.e., 
\begin{equation} 
 p_i^{\rm active}(t,\tau) =   
1 \quad {\rm if} \quad 
\sum_{\tau} |R_i(t+\tau)| \geq L_{\rm Basle} \cdot W_i(t)
\label{prob_act2}
\end{equation} 
Presently, the actual Basle parameter $L_{\rm Basle}$ is  $12.5$, 
but is an open parameter in the model.   
Note, that the restriction to act if 
$\sum_{\tau} |R_i(t+\tau)| \geq L_{\rm Basle} \cdot W_i(t)$
only enforces the agent to adopt the active strategy, i.e., 
to place an (for himself risk-reducing) order on the market. It is
possible, however, that the order is not executed 
because there is no counterparty willing or able
to accept the bet. Under this scenario the actual 
risk of the bank will exceed its regulatory
boundary. This scenario is thus related to a 
(at least local)  banking crisis where due to a lack
of liquidity in the market banks are not able to 
reduce the risk of their positions even when they
are forced to by regulators. This feature of our 
model allows to capture  propagation of
illiquidity effects comparable to what happened 
during the Russian crisis in 1998.

Note, that in the setting described above the 
spread is exogenous, i.e., the agents are not
allowed to adapt the  ''prices'' $\delta$ of the bets. This 
feature of the model makes only sense when
examining pronounced market crashes where empirical 
evidence shows that the adaption of
prices has little or no influence on the willingness 
of agents to trade. However, our framework
can be easily extended to a dynamic endogenous determination of 
the spread variable $\delta$ at each decision stage
at the cost of additional computation burden.

Each of the $N$ players participating in risk trading is represented 
by a node or site. Sites are connected  by links,  
which are non-zero  entries of the 
interaction-graph matrix, $G_{ij}$. A value of $G_{ij}=1$ means site $i$ has 
a connection to site $j$, $G_{ij}=0$ means no connection. 
Players which are connected by links are called neighbors and 
can interact with each other.  In the 
following we shall study the particular classes of different 
network topologies shown in Fig. \ref{Fig_topology}.  
The connectivity of these graphs, defined as the average number of 
links per site 
$C=\langle \frac{ {\rm \# links}}{{\rm  site }} \rangle$,  
models the efficiency of ''information flow'' in the network, a low 
connectivity $C$ means high ''market friction'', i.e., 
not all the deals which would have been possible and 
desirable in the network can be executed due to lack of 
connections. 

The actual model dynamics is shown schematically in Fig. \ref{Fig_flow}. 
At the beginning of each time-step, the external bets are 
offered to the banks. Each bet carries a time of maturity
from now ($t$) till $\tau$ time-steps in the future. As a maximum 
future date for bets we chose $\tau^{\rm max}=10$ time-steps.  
The total exogenous risk entering at this stage at time $t$  is 
$R^{\rm exo}_{\rm total}(t+\tau)= \sum_{i} B^{exo}_i(t+\tau) $. 
After this exogenous update the sites are run through in a random 
asynchronous update. 
Once a site $i$ is chosen, the routine goes through all 
of its neighbor sites $j$ (determined by $G_{ij}$), 
again in  randomized order. 
The actual game between the pair $(i,j)$ now takes place: 

Bank $i$ checks in the market book if any bets are available from 
bank $j$ for all times $\tau$ which can reduce $i$'s risk, i.e., 
$\left|B^{exo}_i(t+\tau)+ \sum_j B_{ij}(t+\tau)+M_{j}(t+\tau) \right| 
<  R_i(t+\tau)$. 
If that is possible the deal will be accepted and $i$ will 
receive the spread $\delta$ from $j$. 
The betting book is now adjusted according to the terms of the 
deal, i.e.,  $B_{ij}(t+\tau)$ now becomes 
$B_{ij}(t+\tau) + M_{j}(t+\tau)$.   
If no more passive strategies are possible $i$ decides 
-- according to Eq. (\ref{prob_act}) --  
if it will issue an offer itself by placing  an entry in the 
market book of size  
$M_i(t+\tau)=R_i(t+\tau)$. If an other bank will (later) 
accept this bet, $i$ will then pay the corresponding spread 
$\delta$ and be then free of risk. 
After all sites have been run through,  all the bets of 
today ($t$) will  be settled according to the random outcome 
of $X(t)$. The wealth update now is nothing but, 
\begin{equation} 
W_i(t+1)=W_i(t) + X_i(t) \cdot 
\left( B^{\rm exo}(t) + \sum_j B_{ij}(t) \right) \quad ,   
\end{equation} 
since all spreads have 
been taken care of along the way. 

A bank is said to be defaulted if its wealth falls below zero. 
In such an event the bank is put into ''receivership''. 
For the remaining duration of the  simulation 
the defaulted bank is not allowed to enter new bets anymore. 
Positive payoffs of  existing bets are collected by the receivership 
whereas negative payoffs are not paid out. 
Any open payments of the defaulted bank to external customers 
will be shared by all the remaining banks. This resembles 
to some extent the function of a deposit insurance system.  
We are aware of the approximative nature of treating the  defaults. 
Numerical experiments however, show that the influence of different 
ways of modeling default events have little impact  
on the results of this study.

A default of a neighbor site can be disastrous for 
a bank, since it can become rapidly exposed to external risk, 
which can be high, and in case of a bad outcome of $X(t)$ can cause its own
default as well. In this context spatial catastrophe-spreading 
can be studied dynamically.

\section{Results}
We implemented the above model for $N=36$ sites to stay within 
reasonable computing times. For simulations of regulation effects 
we used also $N=100$ banks for networks with low connectivity $C$. 
The actual topologies we used for simulations are schematically 
shown in Fig. \ref{Fig_topology}. 
The external spread was set $\delta^{\rm exo}=0$,  
in order to keep things as transparent as possible and not be disturbed by 
externally enforced drifts in the wealth processes. 
As the initial conditions we set $W_i(0)=1$, and 
$R_i(0+\tau)=B_{ij}(0+\tau)=M_i(0+\tau)=0$ for all $i$ and $\tau$.

\subsection{Network Effects}
In order to estimate the effects of different network topologies 
on the game, in a first set of simulations we set the Basle parameter 
$L_{\rm Basle}$ to infinity, i.e., we simulate an unregulated economy
where $L_{\rm Basle}$ does not play any role. 
We first analyze the time-series of the wealth processes.  
In Fig. \ref{Fig_wealth} (a) and (b) we show the wealth trajectories 
of the 36 sites over 200 time-steps 
for an exogenous volatility $D=0.07$. The latter was chosen to avoid 
defaults during this run. The $C=2$ lattice 
(left row) is compared to the fully connected graph $C=35$ (right). 
The difference due to network topology is visible by plain eye. The 
fully connected graph  appears to keep the sites at more or 
less the same level of wealth, while for the badly connected topology 
some sites become ''rich'' and some are at the edge of defaulting. 
Note, that total wealth in the system is not conserved and constant 
but  fluctuates due to the payoffs of the exogenous bets of the clients.
The fully connected graph provides a ''fair'' basis for all sites. 
In Fig. \ref{Fig_wealth} (c) and (d) the log-return of the wealth process,  
$r_i(t) = {\rm log}W_i(t)-{\rm log}W_i(t-1)$, is shown, again 
for all sites.  It can be seen that for the fully connected net 
the average spread 
in returns is significantly smaller, the market is less  
volatile. We now turn to finite values of  $L_{\rm Basle}$. 

\subsection{Regulation Parameter  Effects}
The log-return is frequently used in financial time-series, and 
-- for efficient and complete markets (which is unrealistic) -- 
is often assumed to be Gaussian \cite{Fama65}. 
Log-return  distributions for two different topologies
($C=2$ and $C=35$) are shown in Fig. \ref{ret_distr} (top) 
 for $L_{\rm Basle}=0.05$, which corresponds to a highly 
regulated economy. 
Probabilities (normalized histograms) 
for large events (high risk) are  larger for the $C=2$
network. 
The log-return distribution of the fully connected 
network clearly does not extend as far as the $C=2$. 
In Fig.  \ref{ret_distr} (bottom) the situation for 
the $C=35$ graph is depicted for two different regulation
regimes, $L_{\rm Basle}=0.05$ (highly regulated), and 
$L_{\rm Basle}=5$ (practically unregulated). 
As expected, the probabilities for large log-return is 
higher for the unregulated scenario. 
Common to all the distributions is that they are 
compatible with power law tails. To become more 
quantitative we characterize the distributions 
by power-fits to their tails, the value of the 
probability-density at $r=1$, and the maximum of $r$, which 
occurred in simulations. 
The data is collected in Table 1. 
Fat tails are a well known phenomenon in financial time-series 
\cite{Mandelbrot63,Mueller90,Ding93,Bollerslev94}, 
which has been addressed  in a number of  
physical models recently, see e.g.  
\cite{Farmer99,Iori99,Maslov99,Equiluz00,Jefferies01}.  

In Fig. \ref{Fig_res} a we present the number of defaults  as a function 
of connectivity, occurred during a simulation with exogenous volatility 
$D=0.15$.  All the data and errors shown in the figure are 
mean values  and standard deviations from 100 independent simulation runs, 
each a 200 time-steps long.  
In the highly regulated scheme
($L_{\rm Basle}=0.05$) the topology effect is small,  
in the unregulated world, however, 
a large $C$ has a mayor impact on reducing the  number of defaults. 

It seems reasonable to define an index of 
network efficiency 
which relates the total exogenously imposed risk 
$R^{\rm exo}_{\rm total}(t)$
(at the beginning of a trading day ) 
to the total risk the banks are left with after risk-trading 
(at the end of a trading day): 
\begin{equation}
E(t)= \frac{R^{\rm exo}_{\rm total}(t)
-\sum_{i,\tau}^{N,\tau^{\rm max}} R_i(t+\tau,~{\rm after~trades}) }
{R^{\rm exo}_{\rm total}(t) } \quad . 
\end{equation}
If all the risk is compensated $E=1$,  
if trading did not have any risk-reducing effect, $E=0$. 
The result as a function of $C$ is shown in Fig. \ref{Fig_res} (b). 
For both regulation schemes there is about a 30 percent increase 
in $E$ from $C=2$ to $C=35$. 
System volatility  is defined as 
\begin{equation}
V=\frac{1}{N}\sum_{i,t} \left| r_i(t) \right| \quad , 
\end{equation}
and is a measure of trading activity in  markets.
In the regulated scenario volatility is always below the 
unregulated (c).    
The spread $S$, defined as the variance of wealth of the $N$ sites 
at the latest time in the simulation (200 time-steps), is intended 
as an indicator of system ''un-fairness'',  
and is shown in Fig. \ref{Fig_res} (d).

In Fig. \ref{Fig_res} (e) we show the ratio of active to passive strategies 
as adopted by the players during the run. 
Here ''act'' means the number of active offers placed on the market, 
which were not enforced by the regulation threshold. 
It is seen that for the poorly connected topologies sites  are 
are more active since they face high risk/wealth ratios. 
The ratio is smaller for the highly regulated scheme 
$L_{\rm Basle}=0.05$. ''act'' is not to be confused with 
''\# enforced'' which is the number of enforced trades due 
to  the Basle threshold, i.e., through Eq. (\ref{prob_act2}). 
The situation for the two values of $L_{\rm Basle}$ are shown 
in  Fig. \ref{Fig_res} (f). For the unregulated scheme there 
are no such offers, whereas in the regulated world the number 
of enforced offers is high and declines somewhat for 
higher connectivities. 

\subsection{Global Safety}
As a measure for overall system safety in the various 
networks we looked at the 
mean first-default-times in separate runs for various values 
of the exogenous volatility $D$. The mean first-default-times 
are averages over the time-steps in the simulation until 
the fist default occurs. 
1000 independent runs have been performed; random graphs 
have been additionally averaged over 10 different random topologies. 
For little external risk (small $D$) an approximate power law is seen 
whereas higher external risk levels seem to indicate a 
faster increase of safety with $C$, 
Fig. \ref{Fig_tot}.  

For a fixed value of $D$ we run simulations for different 
values of $L_{\rm Basle}$. 
To study the effects of network type we perform runs 
for all connectivities in a $N=36$ bank world, and 
for the $C=2$  and $C=4$ cases  for $N=100$. 
For 100 banks  higher connectivity was beyond reasonable computation time.  
Results are shown  in Fig. \ref{Fig_basle}, for $N=36$ (top) 
and $N=100$ (bottom).   

As expected, for all connectivities  
network safety becomes  higher as $L_{\rm Basle}$ lowers. 
We find an existence of 2 safety-plateaus,  which 
are separated by a phase transition like drop. 
This suggests a  
strong connection to statistical physics: 
The regulation parameter $L_{\rm Basle}$ plays the role of a temperature, 
the mean first-default-time becomes the order parameter.  
In the small $L_{\rm Basle}$ limit 
all banks are forced to be in the active state
and the system approaches an ''ordered'' phase in terms of strategies. 
The origin of the plateaus is not entirely surprising: 
 $L_{\rm Basle}$  enters the system as a threshold 
upon which trading is enforced. As this threshold is lowered 
the number of forced offers (''\#forced'') rises. This is demonstrated in 
Fig.  \ref{Fig_stat} (a). For $C=35$ (630 links) the absolute number of 
forced offers is still less than for $C=2$ (36 links)
for all $L_{\rm Basle}$. 
The number of active offers ''\#act'' (not enforced by 
$R_i(t+\tau)/W_i(t)\geq L_{\rm Basle}$) is more or less 
independent of $L_{\rm Basle}$ for $C=2$, and increases slightly 
for $C=35$ converging to the level of $C=2$ in the unregulated limit (b). 
The number of passive trades (c) follows the number of offers 
(\#forced  + \#act). It is seen that in the highly regulated regime 
there is always more offers than passive trades, i.e., an 
oversupply of offers. Trading activity saturates at this point 
since no more passive trades are needed to compensate 
external risk and to reach locally optimal risk levels.  
This explains the existence of the left plateau.  
The right plateau in the unregulated regime occurs when 
the number of offers  and passive trades approach an equilibrium (same 
levels). 
The situation looks similar for both lattice sizes.  
Naively comparing 
figures \ref{Fig_basle} (top) and (bottom) 
for $C=2$ and $C=4$ suggests that the 
smaller network is slightly safer. This is of course not so if 
one considers the individual risk of default within those networks. 
A further observation is that in the small $L_{\rm Basle}$ limit
safety levels are separated equidistantly (in log units) 
for the equidistantly separated  (in log units) connectivities. 
This indicates a power law of safety as a function of $C$ 
in this regime. 
Finally, for completeness we give the ratio of 
active offers to passive trades in Fig. \ref{Fig_stat} (d).

The strong decay from one plateau to the 
other is often related to a phase transition in physics. 
This means the existence of two stable regimes (where 
regulatory activity has no effect) 
which can change from one state to the other within a relatively 
small critical region. Whenever a phase transition is 
present powerful  mathematical tools from statistical 
physics -- such as the existence of universality classes -- 
can be applied and used to describe, and 
more importantly, relate models to real-world systems.

\subsection{Effects of Cyclicality}
We study how cyclicality in inter-bank connections
effect default probabilities or global safety. 
For this end we fix the number of nodes
and links , i.e. the connectivity $C$, 
and vary the number of cycles of given lengths.
We focus especially on small cycles composed of 
3 links (3-cycles) which can be used 
to define the clustering coefficient 
$CC=\frac{3 \,\,\times (\# {\rm \,\,3-cycles})}
{\# \,\, {\rm  connected \,\, node-tripples}}$ of networks
\cite{Newman01}. 
The studied graphs are shown in Fig. \ref{Fig_cycl}. 
The graphs differ in the spectrum of cycles. 
The number of 3-cycles decreases from 
36, 18, 9, and 0 in Figs. \ref{Fig_cycl} 
a, b, c, and d, respectively.  
The mean values and standard deviations 
of first default 
times of these networks based on 100 independent simulations per 
graph are contained in Tab. \ref{Tab_cycl}. 
For the regulated region ($L=0.05$) an increase in 
safety as a consequence of less 3-cycles can be 
inferred, for the unregulated ($L=5$) case such 
a trend is not visible, given the size of errors.  
This indicates that the influence of cyclicality 
increases with the degree of regulation. 
This result implies that in regulated regimes 
banking systems with risk-sharing cooperations, 
schematically represented by  Fig. \ref{Fig_cycl} a,  
have higher global risk than systems with lower cyclicality. 

\subsection{Contagion Effects}
In a set of simulations we observed contagion effects.
Contagion means that the default of a given bank 
significantly increases  the default probability   
of its neighbors. In many scenarios the default of 
neighbors is realized and local ''banking crises'' 
can be observed.  
To model effects like September 11 we artificially 
removed a single large (wealthy) bank from the network at a 
given simulation time-step. 
The spread of the crisis depends heavily on the 
network topology, and ranges from small, locally 
bounded events to the total collapse of the network. 
Since these results are hard to visualize 
over the course of time we do not present them in this 
version of the paper.

\subsection{Small World Effects?}
We moreover looked for topology effects other than connectivity 
and cyclicality alone.
To this end we started from regular networks with a given  $C$ 
and let links ''diffuse'' in a way that kept the graphs connected. 
We could such ''heat'' the graphs from  regular towards  random 
structure with the same $C$ in the spirit of small-world networks 
\cite{Strogatz01}. 
Results from these studies showed 
little influence on ''heating'' time and it is save to 
conclude that for the presented measures the relevant topology 
parameter is $C$.    

\section{Discussion}
To summarize, we have introduced a relatively simple model 
of interacting agents who change  their modes of interaction 
(trading strategy) according to their  state of being, i.e., 
their need to reduce risk.  
The basic model is inspired from the  structure of the well-known 
iterated prisoner's dilemma 
\cite{Axelrod84,Sigmund98}, where two players 
are equipped with two possible actions, see also \cite{Weibull96}. 
In our case the pay-off 
matrix additionally depends on an external process, and the size of 
bets, which in our model are dynamical variables.

The present model contains three relevant parameters 
$C$, $D$, and $L_{\rm Basle}$.  We have checked that the 
remaining parameters, the average riskiness of agents $\alpha$, 
and the maximum time horizon of 
financial instruments $\tau^{\rm max}$, are comparably 
irrelevant. These parameters can be used for scaling the model 
to realistic data. 
In reality the bid-ask-spread $\delta$ is a quantity which crucially depends
on the supply and demand of bets. If banks need to reduce risk 
and can not find counterparties who accept their offers for 
a certain $\delta$, they will have to increase the incentive $\delta$. 
We have so far run the simulations for two additional   
values of $\delta$ (5 \% and 20 \%), 
fixed for all bets. With lower values of $\delta$ we 
see a tiny increase in trading activity. This can be understood 
since active strategies become more attractive, 
see Eq. (\ref{prob_act}),  which then 
also implies slightly more passive trades. 
This underpins the importance 
to incorporate some sort dynamic  price formation in the model,   
i.e., to make $\delta$ a dynamical variable. 
We plan to include this feature in future work.

We have performed several runs 
with various sizes of $N$ to check for finite size effects. 
We found that for $N$ smaller than about 20, scaling starts 
to vanish.  
Our main results are that in a number of crucial measures there exists a 
very sensitive dependence on the connectivity 
(which models the market structure in the real world). 
We find that in  well connected  networks little spread of 
wealth is able to develop  and that topology alone can lead to 
considerably ''fairer'' networks. 
Highly connected networks show significantly 
less large moves in wealth changes (less volatility), the market 
becomes less hectic. Distributions of the log-returns are 
less fat tailed, but still show realistic fat tails, which 
have realistic power-law behavior. 
Well connected networks are more efficient in reducing global risk, 
and show significantly fewer defaults. 
The average first default time (safety) increases with connectivity.  

We finally remark that in reality the parameters $C$,  $D$, and particularly 
$L_{\rm Basle}$ can be  
controlled by central banks and governments to regulate risk 
\cite{Lohmann98}. 
The most interesting  aspect of this paper is the 
existence of two plateaus, one for low and one for high values of 
$L_{\rm Basle}$. This is independent of the network structure. 
It is a major finding that the requirement of a capital 
cushion in the form of the Basle multiplier $L_{\rm Basle}$ as 
currently used by regulators may have unexpected adverse effects.
We found that for some $L_{\rm Basle}$ a   
seemingly strong reduction of the regulatory parameter 
has vanishing or even zero effect on the safety of the system. 
This may lead to unjustified  overconfidence in the regulatory 
action. It may also lead to unnecessary restrictions 
to economic activity, because a relaxation of 
the Basle multiplier would not lower the safety of the system. 
Moreover, we show that in regulated regimes 
banking systems with risk-sharing cooperations 
have higher global risk than systems with lower cyclicality. 

The next stage of the analysis should focus on a more detailed 
inspection of the interrelation of the model parameters and 
on the calibration of the model to real world data. 
A further promising extension bringing the model closer to 
reality will be the inclusion the risk assessment of agents. 
The model is structured in such a way that such an inclusion 
will involve just to make the information of the betting books 
available to all the banks. 

Even though our model is phrased in terms of 
a interactions between banks who insure themselves by trading 
financial assets the core of the model should be very general 
and can be seen as a starting point for a dynamical theory of 
insurance. 
 
\vfill

\noindent
{\bf Acknowledgments:} We thank the SFI and in particular J.D. Farmer 
for their kind hospitality and inspiring atmosphere in April and May 
2002, where parts of this work were written.

\newpage

\newpage 
\begin{table}[htb]
\caption{Characterization of the log-return distributions for 
various values of $N$, $C$, and $L_{\rm Basle}$, with a 
given exogenous volatility $D=0.15$. The tail exponent is the 
power exponent to the tails of the distributions. 
max$(r)$ and $p(.)$ correspond to simulations with 720,000 (2m) 
changes in wealth levels, for the $N=36$ (100) networks.  
}
\begin{center}
\begin{tabular}{c c c c c c }
\hline
$N$  &$ L$  & $C$  & tail exponent & max ($r$) & $p(r=1)$ / $10^{-5}$ \\
\hline
\hline
36   & 0.05 & 2    &  -3.87        & 5.74      & 3.0556  \\
36   & 0.05 & 35   &  -3.79        & 4.27      & 0.8333  \\
36   & 5    & 2    &  -3.71        & 6.53      & 3.4722  \\
36   & 5    & 35   &  -3.30        & 4.99      & 1.9445  \\
\hline
100  & 0.05 & 2    &  -3.87        & 8.10      & 2.7501  \\
100  & 5    & 2    &  -3.62        & 5.54      & 2.9001  \\
\hline
\end{tabular}
\end{center}
\end{table}

\begin{table}[htb]
\caption{Influence of cycle spectra on the 
first default times on graphs with fixed 
number of nodes ($N=36$) and links (63), i.e., fixed global 
connectivity ($C=1.75$). The 
corresponding graphs are shown in Fig. \ref{Fig_cycl}. 
Data corresponds to mean values and standard deviations of 
100 independent simulations per graph.  
For $L=0.05$ an increase in safety as a consequence of less 
3-cycles can be inferred. For the  $L=5$ case such a trend  
is not visible, given the size of errors. 
}
\begin{center}
\begin{tabular}{c c c c c }
\hline
$L$  &FDT (Fig. \ref{Fig_cycl} a)&FDT (Fig. \ref{Fig_cycl} b)&FDT (Fig. \ref{Fig_cycl} c)&FDT (Fig. \ref{Fig_cycl} d)\\
\hline
\hline
0.05 & 128.31(47.46) & 145.76(65.50) & 161.28(52.52) & 165.99(65.21) \\
5    &  47.26(19.96) &  42.24(16.93) &  41.33(19.59) &  41.97(19.14) \\
\hline
\end{tabular}
\end{center}
\label{Tab_cycl}
\end{table}

\newpage

\begin{figure}[H]
\caption{
Schematic plot of different graph topologies.   
(a) One site is connected to all the other $N-1$ sites (monopoly).  
(b) Each site is connected to two neighboring sites $C=2$ ( 1 dim. circle, 
with periodic boundary) 
(c) Each site is connected to four neighboring sites $C=2$ 
(regular 2 dim. lattice, with p.b.)
(d)  Random lattice: Each site is connected to random other sites, 
the average number of links per site is fixed. ''Random 0.222'' means 
that on average 22.2 \% of all possible ($N\times N$) links are present. 
(e) Same as before with ''Random 0.444'', i.e., 44.4 \% of $N\times N$
possible links are realized on average. 
These numbers are chosen to provide a connectivity of $C=8(16)$ 
links/site on graphs with $N=36$ or $N=100$ sites. 
For the random graphs we checked that they were complete, i.e., 
every site could be reached from every other site. 
(f) Each site is connected to all the other sites $C=35$ 
(fully connected).  
}
\label{Fig_topology}
\end{figure} 

\begin{figure}[H]
\caption{
Flow diagram for the model. The dynamics during a single time-step 
$t$ is shown schematically.  
}
\label{Fig_flow}
\end{figure} 

\begin{figure}[H]
\caption{
Comparison between the $C=2$ (left)  and the fully connected $C=35$ 
(right) topologies.  Top: Wealth trajectories for the 36 sites. 
Bottom: Log-returns $r_i(t)$ of the sites plotted on top of each other. The 
volatility for the fully connected graph is clearly less. 
}
\label{Fig_wealth}
\end{figure} 

\begin{figure}[H]
\caption{
Distribution functions derived from histograms of $r_i(t)$  
for comparison of different connectivities (top) 
and regulation schemes (bottom). Power fits 
(in regions indicated by the lines) 
to the tails are gathered in Table 1. 
}
\label{ret_distr}
\end{figure} 

\begin{figure}[H]
\caption{
Performance measures as a function of the different topologies
at two regulation extremes, $L_{\rm Basle}=0.05$ (highly regulated)
and $L_{\rm Basle}=5$ (practically no regulation).  
(a) Number of defaults occurred within the first 200 
time-steps. 
(b) Efficiency parameter $E$. 
(c) Volatility estimate $V$. 
(d) Spread $S$, a measure for ''un-fairness'' of the system.   
(e) Ratio of active offers (without Basle, 
    Eq. (\ref{prob_act2})) to passive trades. 
(f) Number of forced active offers (by the Basle 
    condition, Eq. (\ref{prob_act2}))
}
\label{Fig_res}
\end{figure} 

\begin{figure}[H]
\caption{
    Mean first-defaults-time  (a measure of system safety) 
    for several values of $D$ for a practically  unregulated 
    economy ($L_{\rm Basle}=5$). 
    Errors are less than double symbolsize. 
}
\label{Fig_tot}
\end{figure} 

\begin{figure}[H]
\caption{
Mean first-default-time (a measure of system safety) as a function of 
the regulation parameter $L_ {\rm Basle}$ for various values of $C$ 
on a $N=36$ (top) and $N=100$ (bottom) lattice at a fixed value of  $D=0.15$. 
Errorbars are about twice symbolsize and are omitted for clarity.}
\label{Fig_basle}
\end{figure} 

\begin{figure}[H]
\caption{
Statistics of the same runs as in the previous plot: 
(a) number of enforced offers (by the Basle condition)  
as a function of $L_ {\rm Basle}$. (b) and (c) show the 
numbers of active offers (without Basle enforcement) 
and passive trades respectively. 
Finally, in (d) the ratio of active 
offers compared to passive trades is shown.}
\label{Fig_stat}
\end{figure} 

\begin{figure}[H]
\caption{
Connection topology for the study of cycle effects. 
In these graphs the number of nodes and links are fixed. 
Different connection topology alters the number of cycles 
of given length in the total graph. The spectrum 
of cycle size shifts towards larger cycles from 
(a) to (d). The number of shortest cycles (length 3) 
is 36, 18, 9, and 0 for Figs. (a), (b), (c), and (d), respectively. 
The number of cycles involving four links increase,  
9, 18, 18, and 27 for (a), (b), (c), and (d). Effects 
of the different cycle spectra are collected in Table
\ref{Tab_cycl}.}   
\label{Fig_cycl}
\end{figure} 

\newpage
\begin{figure}[H]
\begin{center}
\begin{tabular}{ll}
\includegraphics[width=3.5cm]{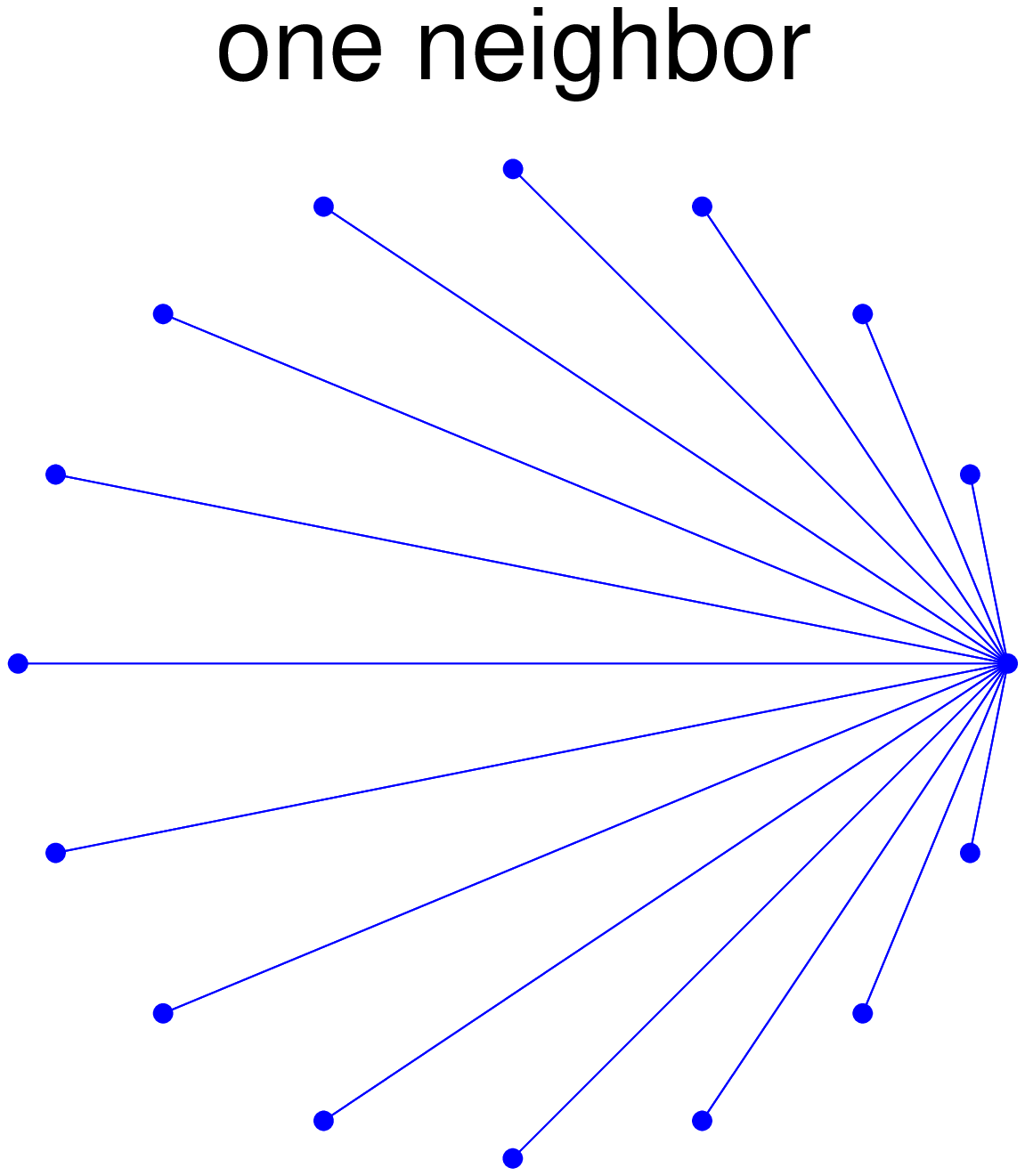} &  
\includegraphics[width=3.5cm]{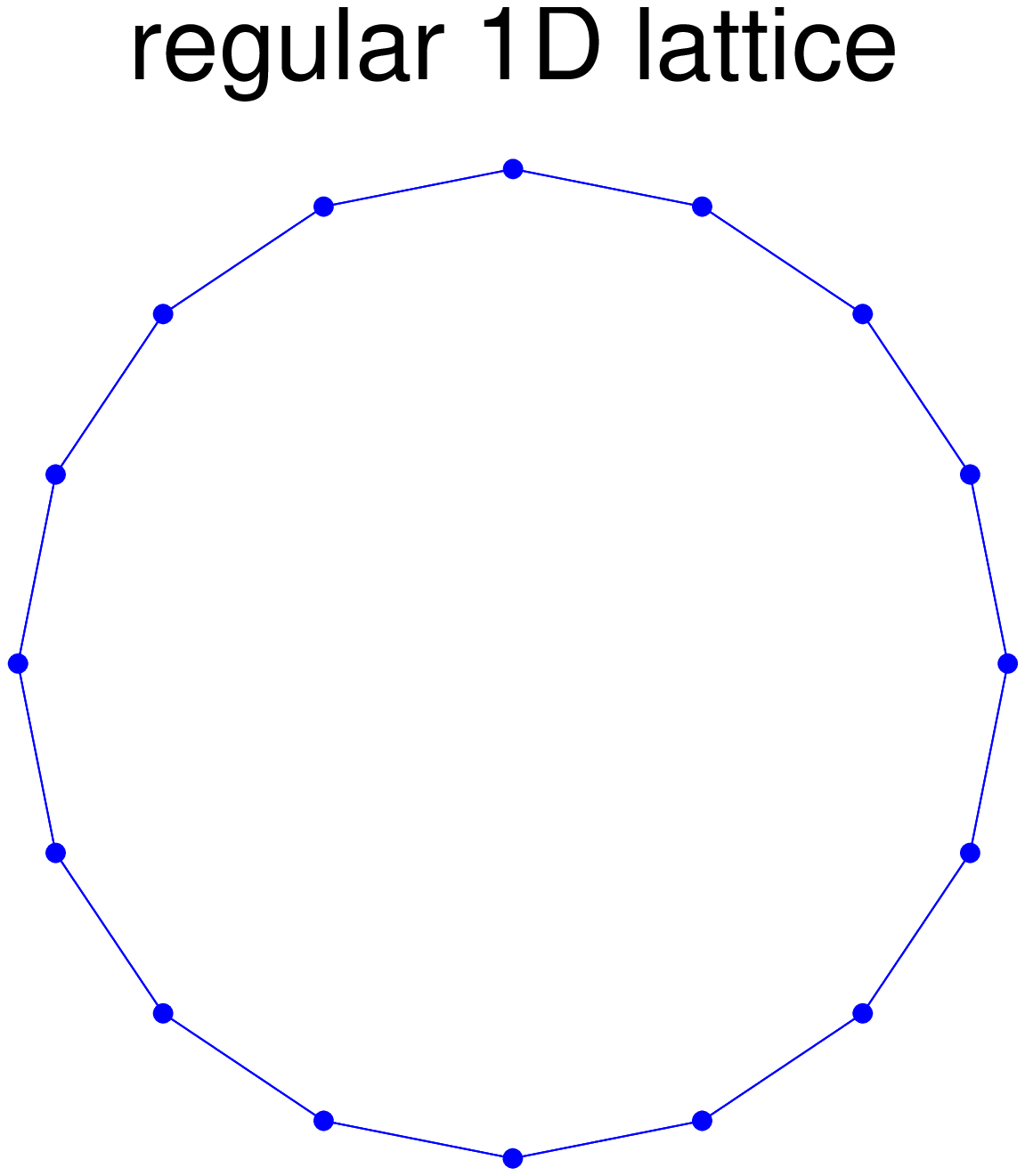} \\ 
 & \\
\includegraphics[width=3.5cm]{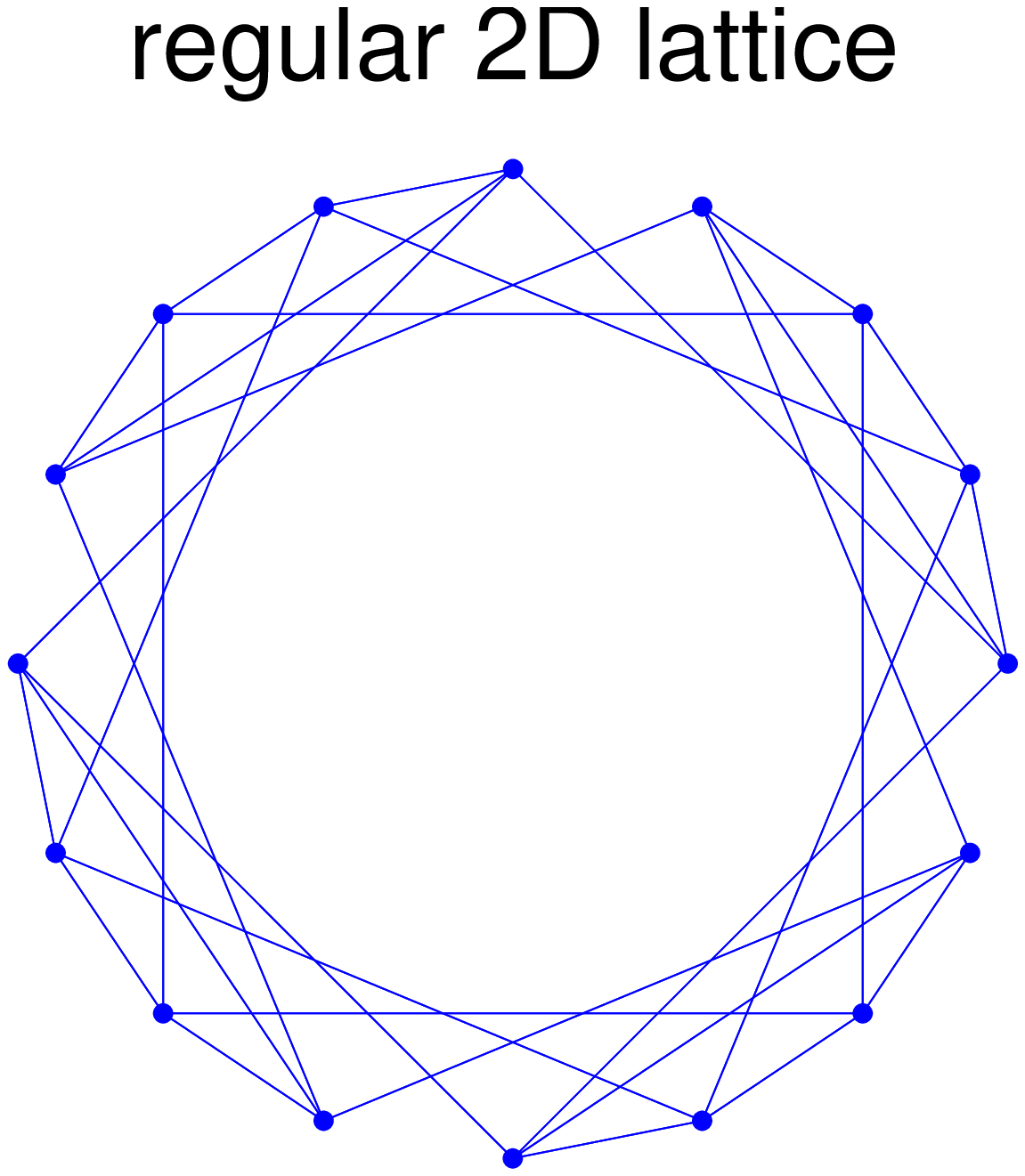} &  
\includegraphics[width=3.5cm]{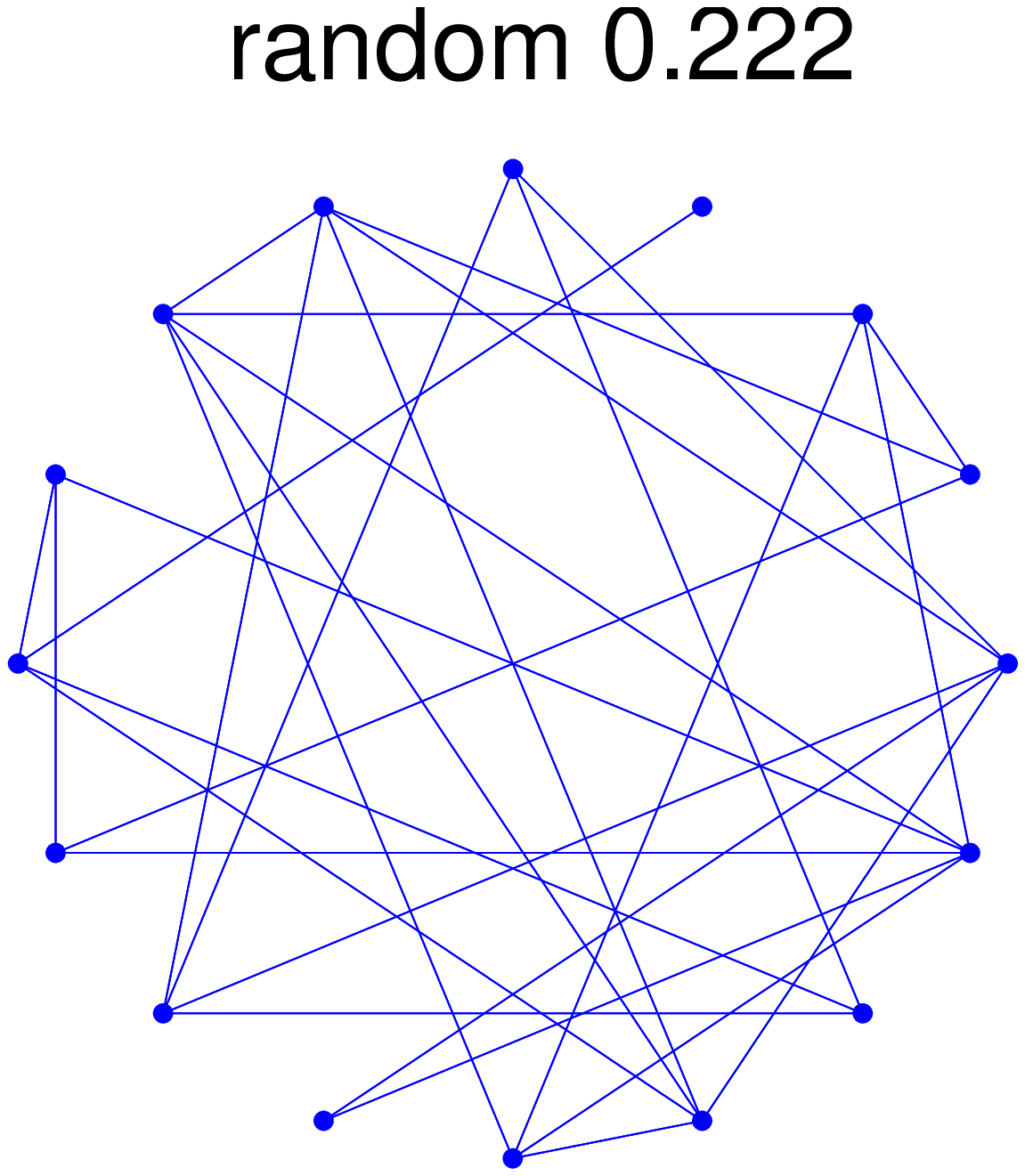} \\ 
 & \\
\includegraphics[width=3.5cm]{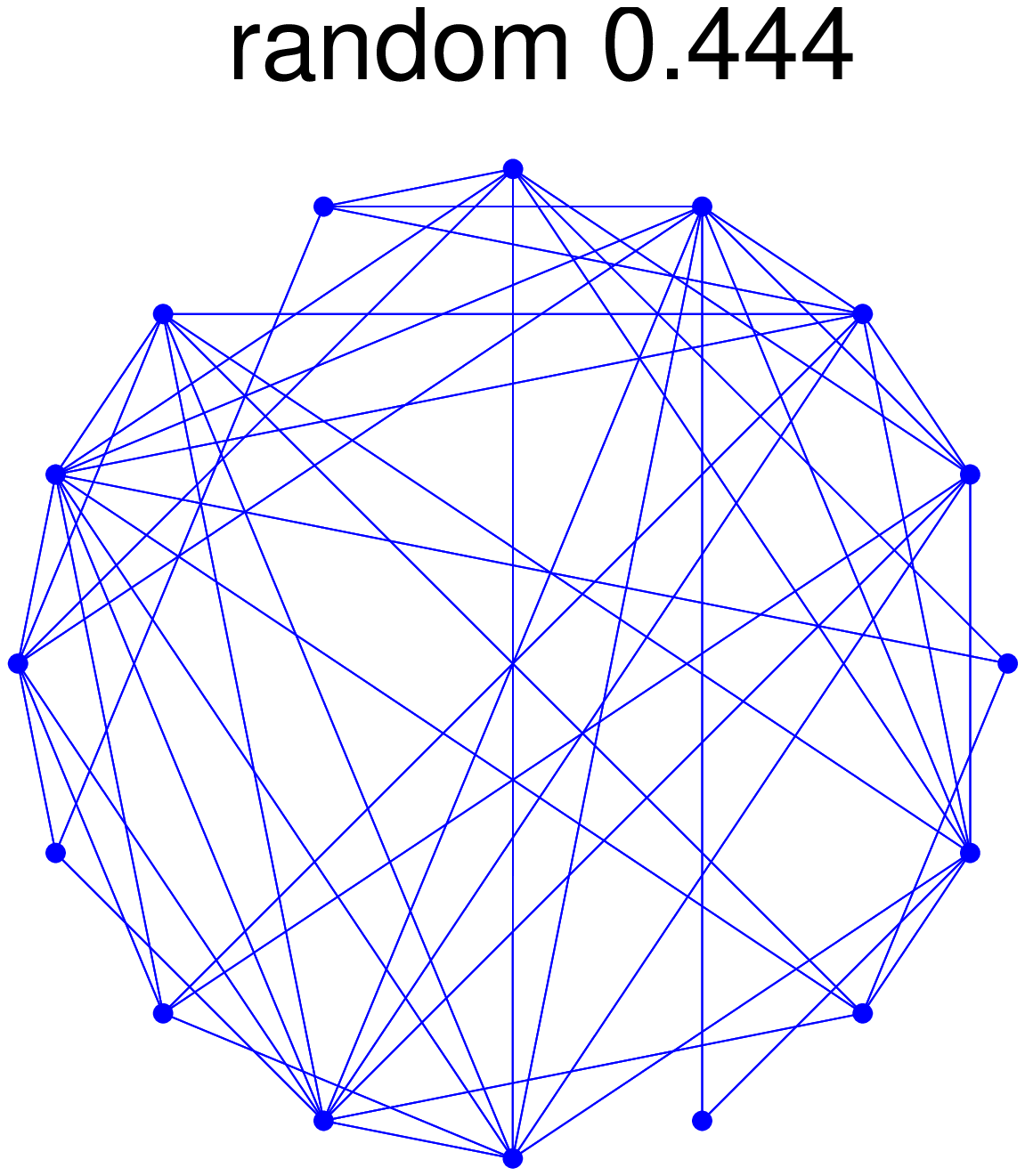} &  
\includegraphics[width=3.5cm]{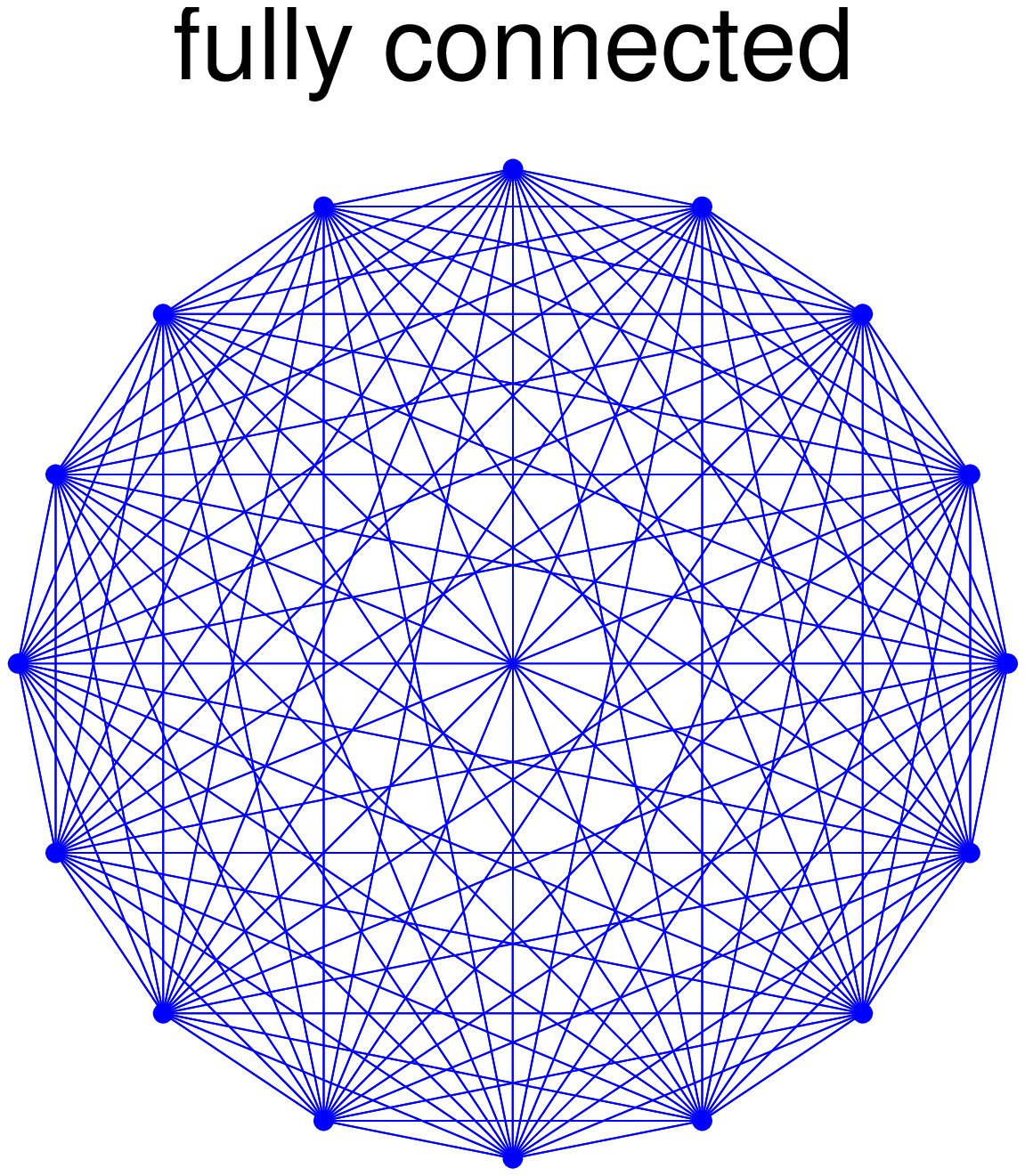} \\ 
\end{tabular}
\end{center} 
\end{figure} 

\vfill
\begin{center}
{\Large FIG. 1}
\end{center} 
\newpage

\begin{figure}[htb]
\begin{center}
\begin{tabular}{ll}
\includegraphics[width=10.5cm]{bank_flow} \\ 
\end{tabular}
\end{center} 
\end{figure} 

\vfill
\begin{center}
{\Large FIG. 2}
\end{center} 
\newpage

\begin{figure}[htb]
\begin{center}
\begin{tabular}{ll}
\includegraphics[width=16.0cm]{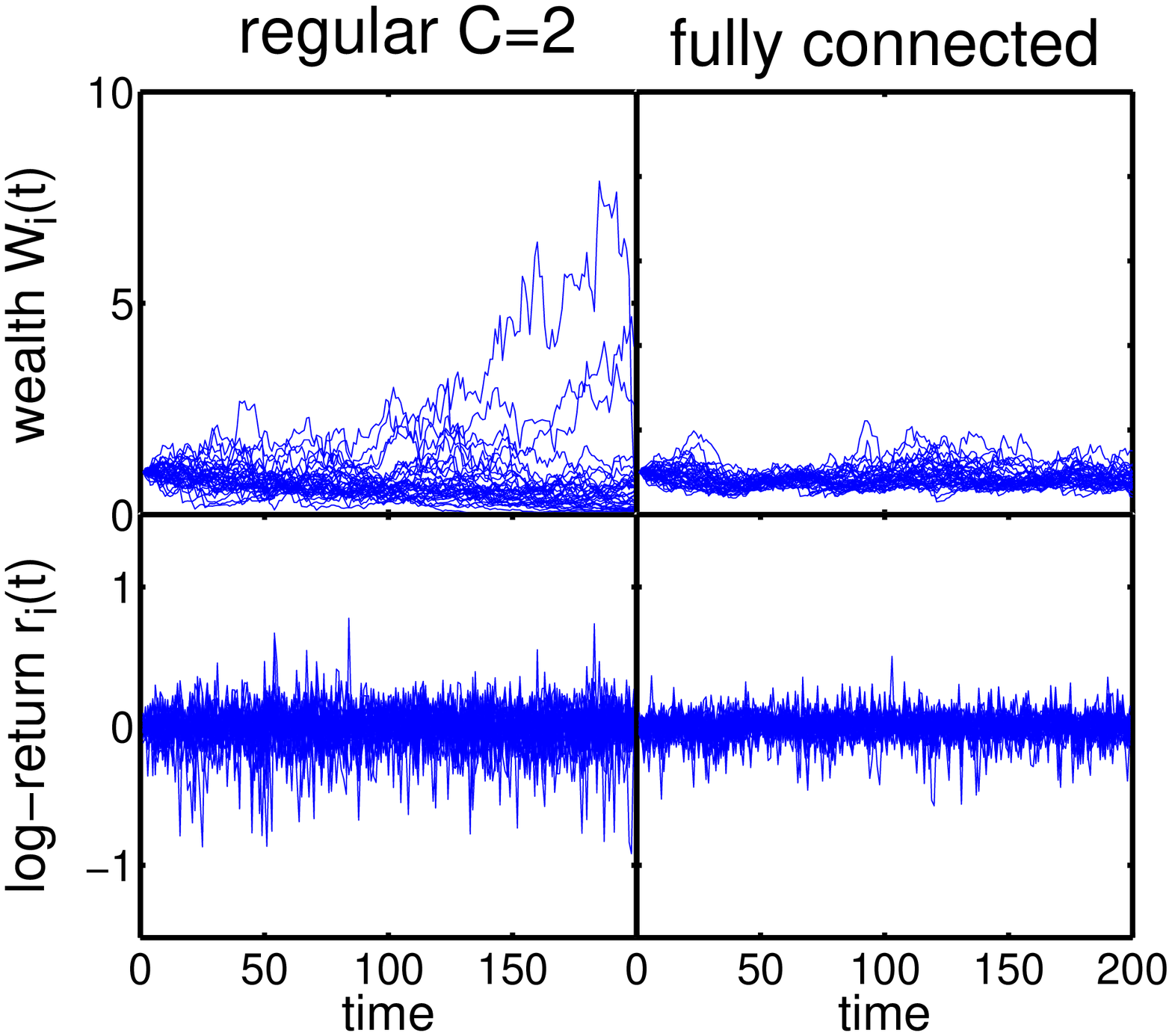} &  \\
\end{tabular}

\vspace{-19.7cm} 
\hspace{5.5cm} {\LARGE (a)}  \hspace{5.5cm} {\LARGE (b)}  \\

\vspace{5.0cm} 
\hspace{5.5cm} {\LARGE (c)}  \hspace{5.5cm} {\LARGE (d)}  \\

\vspace{5.5cm} 

\end{center} 
\end{figure} 

\vfill
\begin{center}
{\Large FIG. 3}
\end{center} 
\newpage

\begin{figure}[htb]
\begin{center}
\begin{tabular}{l}
\includegraphics[width=10.5cm]{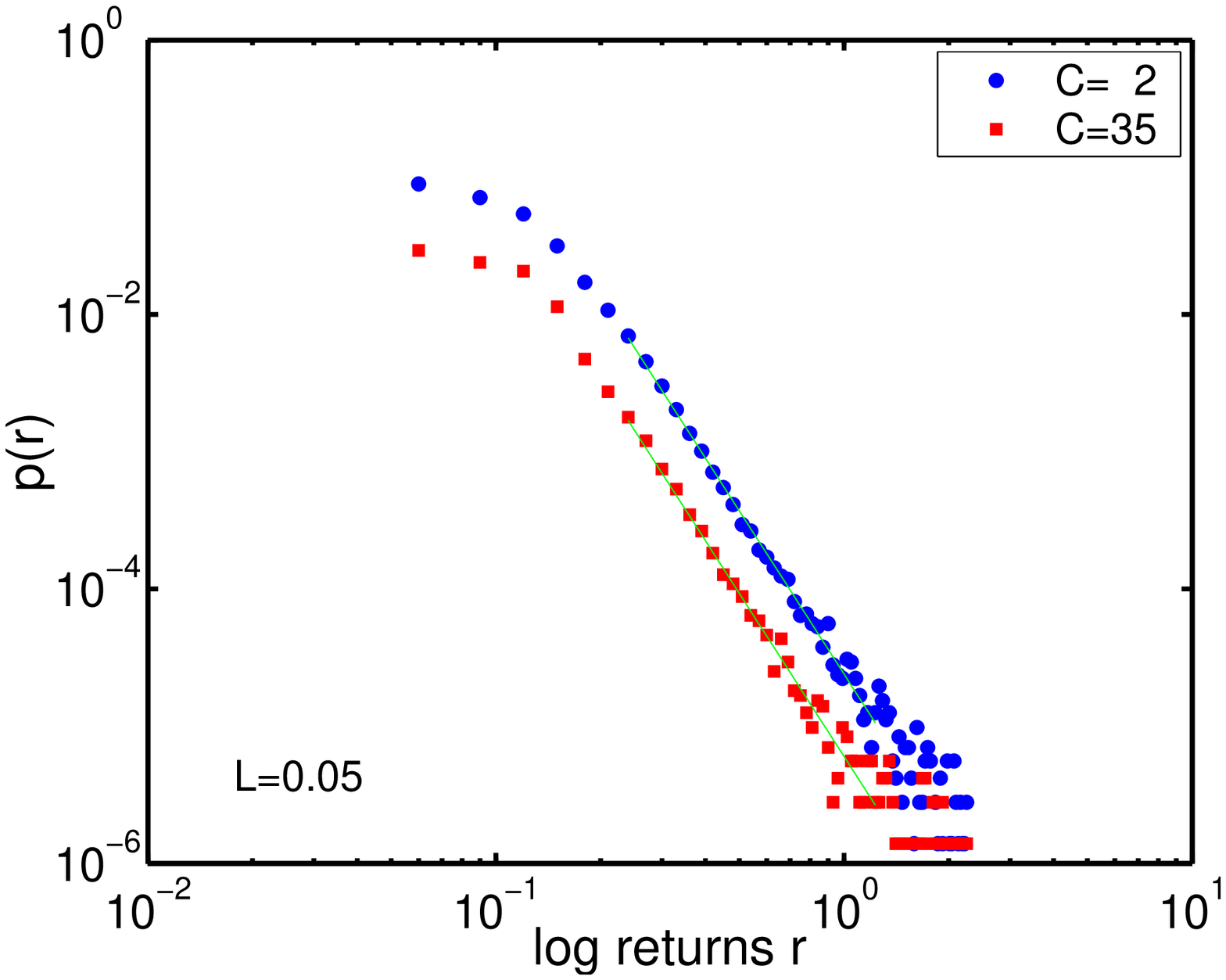} \\
\includegraphics[width=10.5cm]{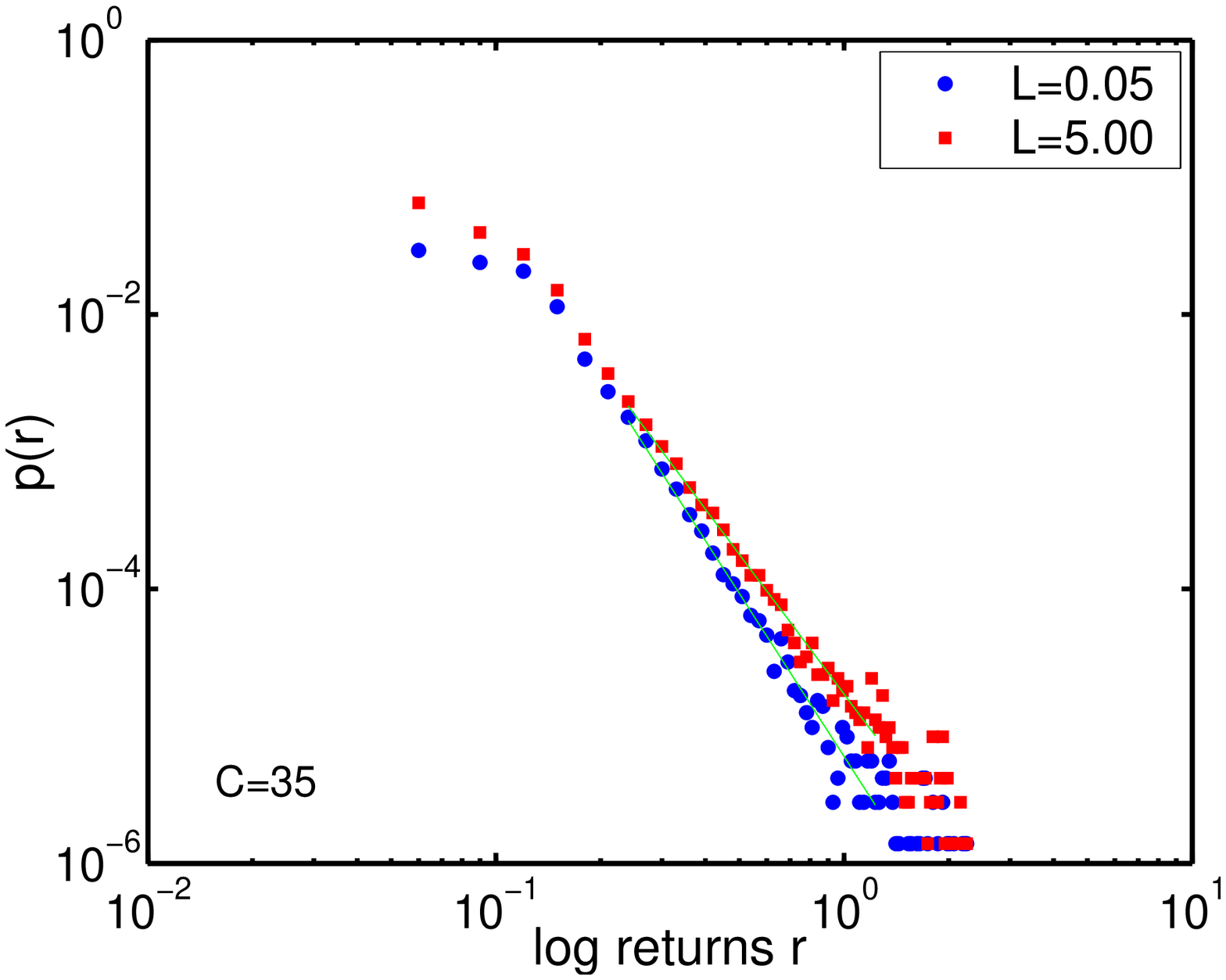} \\
\end{tabular}
\end{center} 
\end{figure} 

\vfill
\begin{center}
{\Large FIG. 4}
\end{center} 
\newpage

\begin{figure}[tb]
\begin{center}

\vspace{1.0cm} 
\begin{tabular}{cc}
\includegraphics[width=8.0cm]{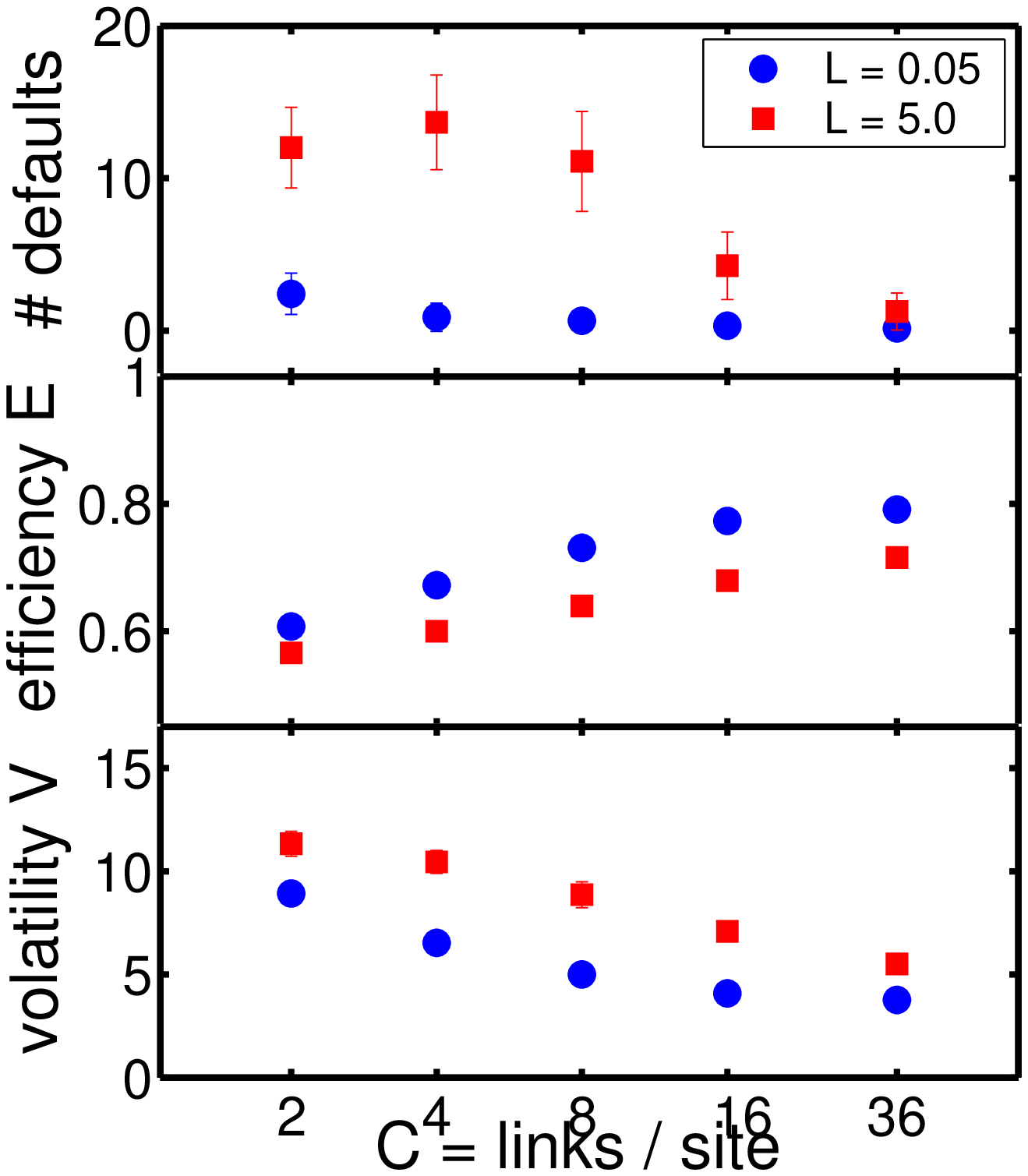} &
\includegraphics[width=8.0cm]{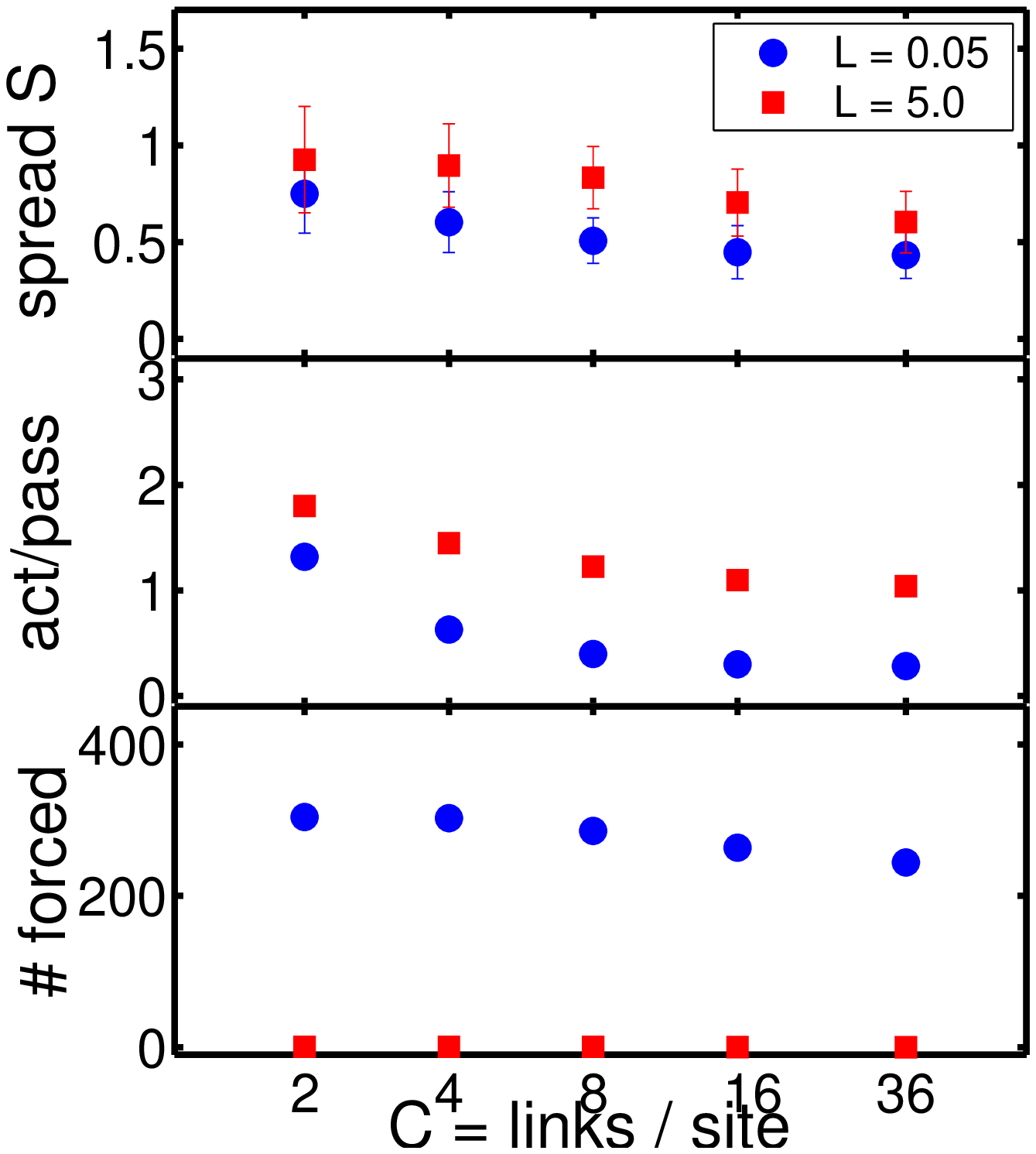} \\
\end{tabular}

\vspace{-7.2cm} 
\hspace{-12.5cm} {\LARGE (a)} \\

\vspace{1.9cm} 
\hspace{-12.5cm} {\LARGE (b)} \\

\vspace{1.9cm} 
\hspace{-12.5cm} {\LARGE (c)} \\

\vspace{-5.9cm} 
\hspace{3.9cm} {\LARGE (d)} \\

\vspace{1.9cm} 
\hspace{3.9cm} {\LARGE (e)} \\

\vspace{1.9cm} 
\hspace{3.9cm} {\LARGE (f)} \\

\vspace{9.7cm} 
\end{center} 
\end{figure}

\begin{center}

\vfill
{\Large FIG. 5}
\end{center} 
\newpage

\begin{figure}[htb]
\begin{center}
\begin{tabular}{c}
\includegraphics[width=14.0cm]{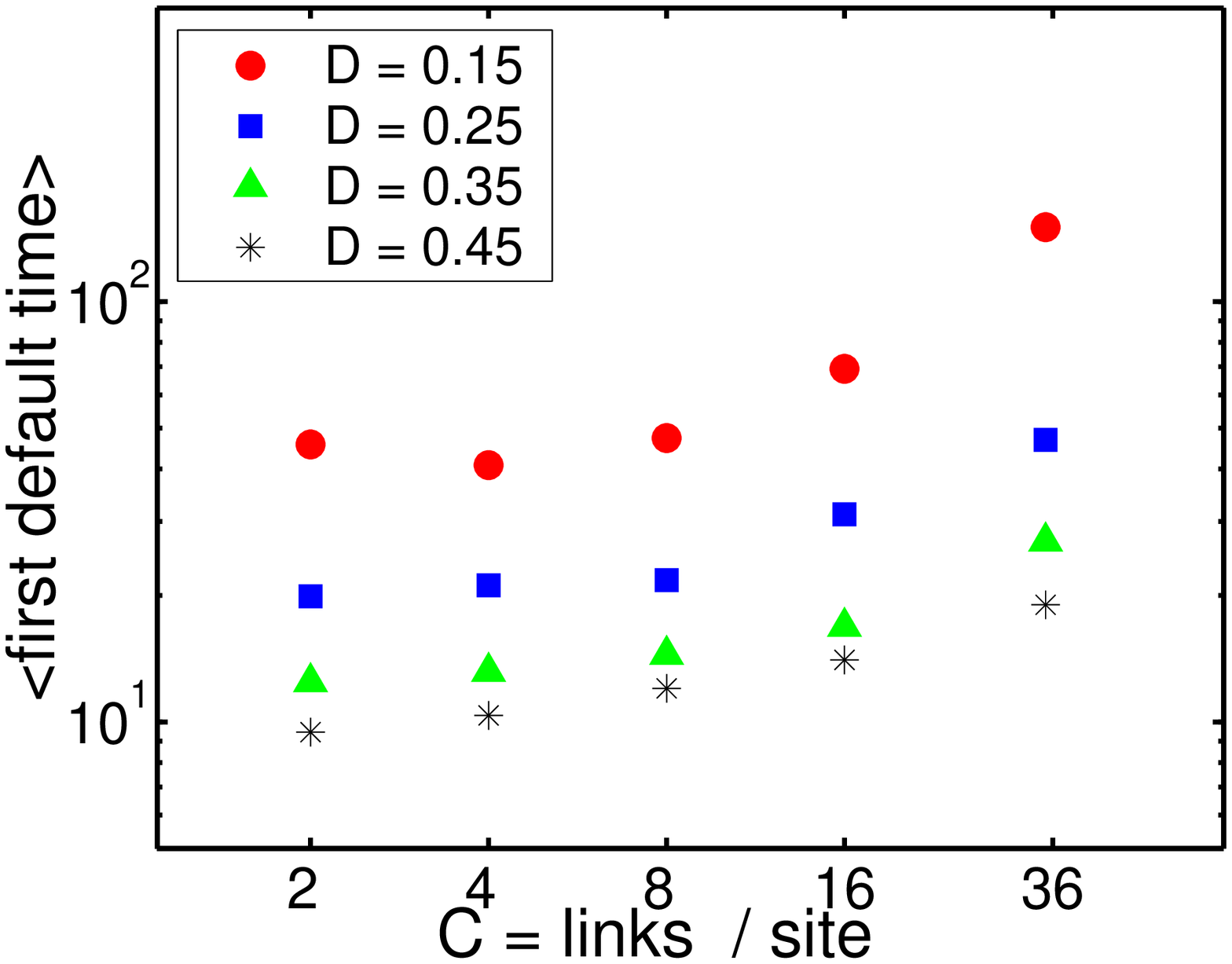} \\   
\end{tabular}
\end{center} 
\end{figure} 

\vfill
\begin{center}
{\Large FIG. 6}
\end{center} 
\newpage

\begin{figure}[htb]
\begin{center}
\begin{tabular}{cc}
\includegraphics[width=11.0cm]{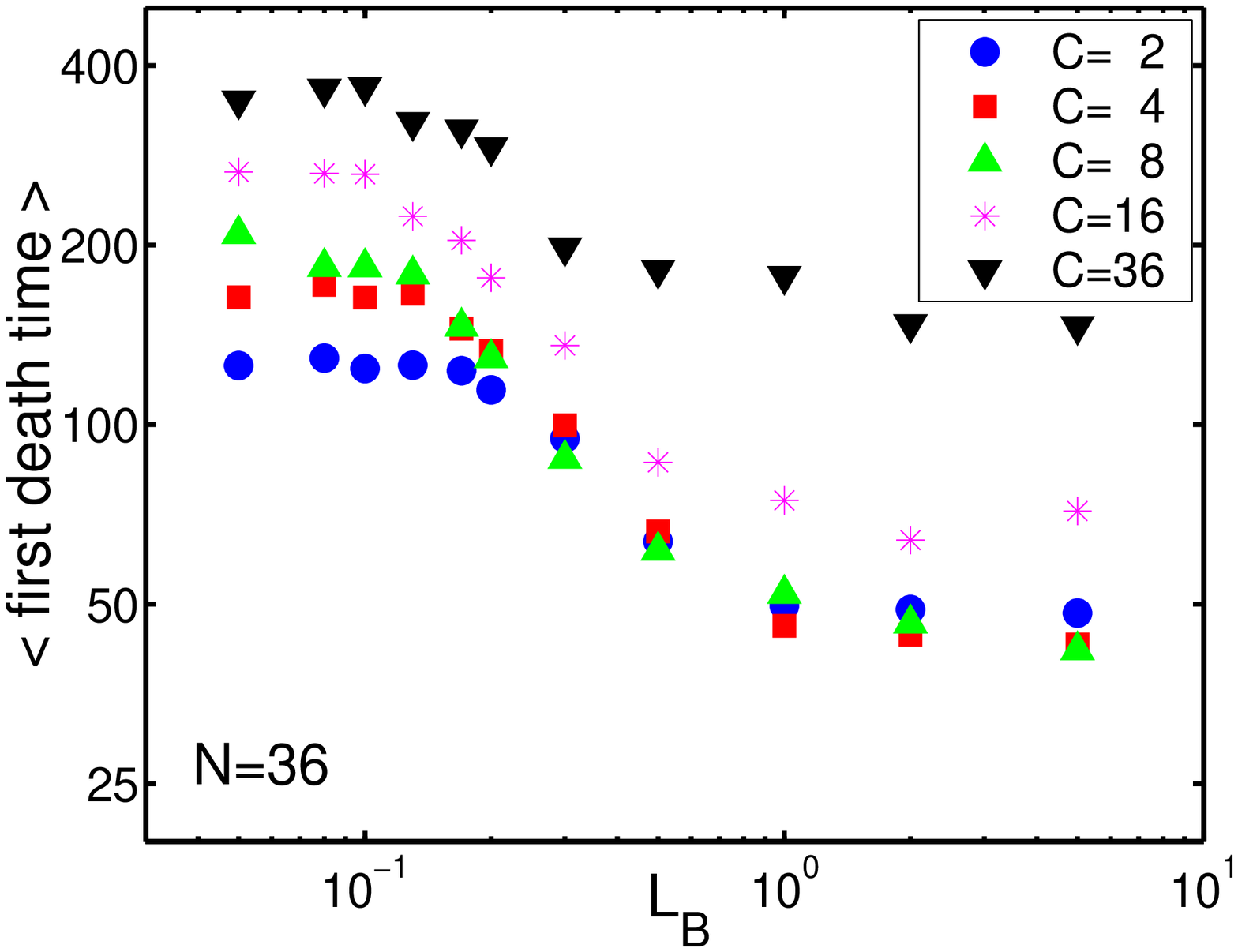} \\
\includegraphics[width=11.0cm]{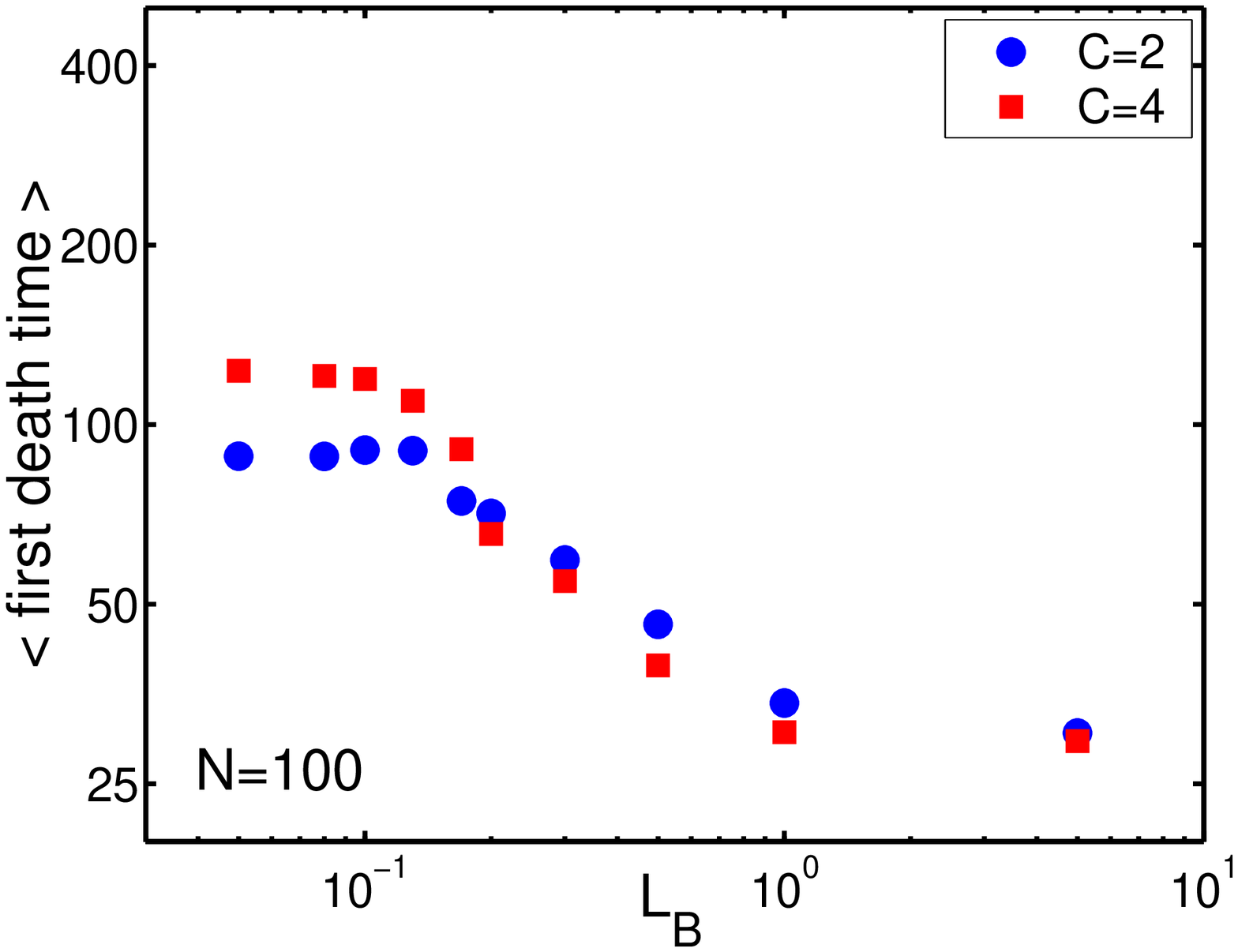} \\
\end{tabular}
\end{center} 
\end{figure} 

\vfill
\begin{center}
{\Large FIG. 7}
\end{center} 
\newpage

\begin{figure}[htb]
\begin{center}
\begin{tabular}{c}
\includegraphics[width=15.9cm]{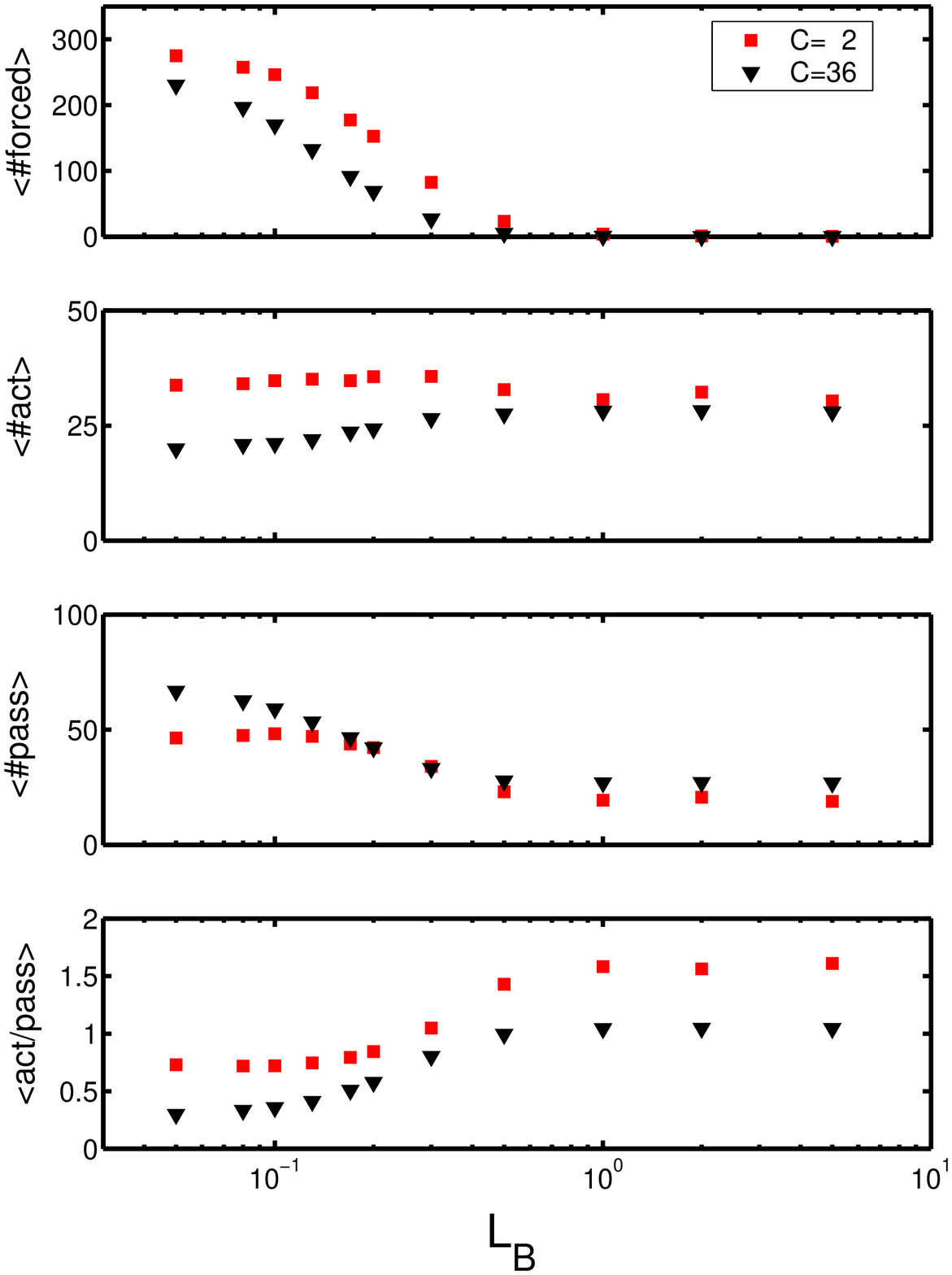} \\   
\end{tabular}
\vspace{-18.7cm} 

\hspace{13.5cm} {\LARGE (a)} \\
\vspace{4.3cm} 
\hspace{13.5cm} {\LARGE (b)} \\
\vspace{4.3cm} 
\hspace{13.5cm} {\LARGE (c)} \\
\vspace{4.3cm} 
\hspace{13.5cm} {\LARGE (d)} \\
\vspace{3.5cm} 
{\Large FIG. 8}
\end{center} 
\end{figure} 
\newpage

\begin{figure}[htb]
\begin{center}
\begin{tabular}{cc}
\includegraphics[width=4.0cm]{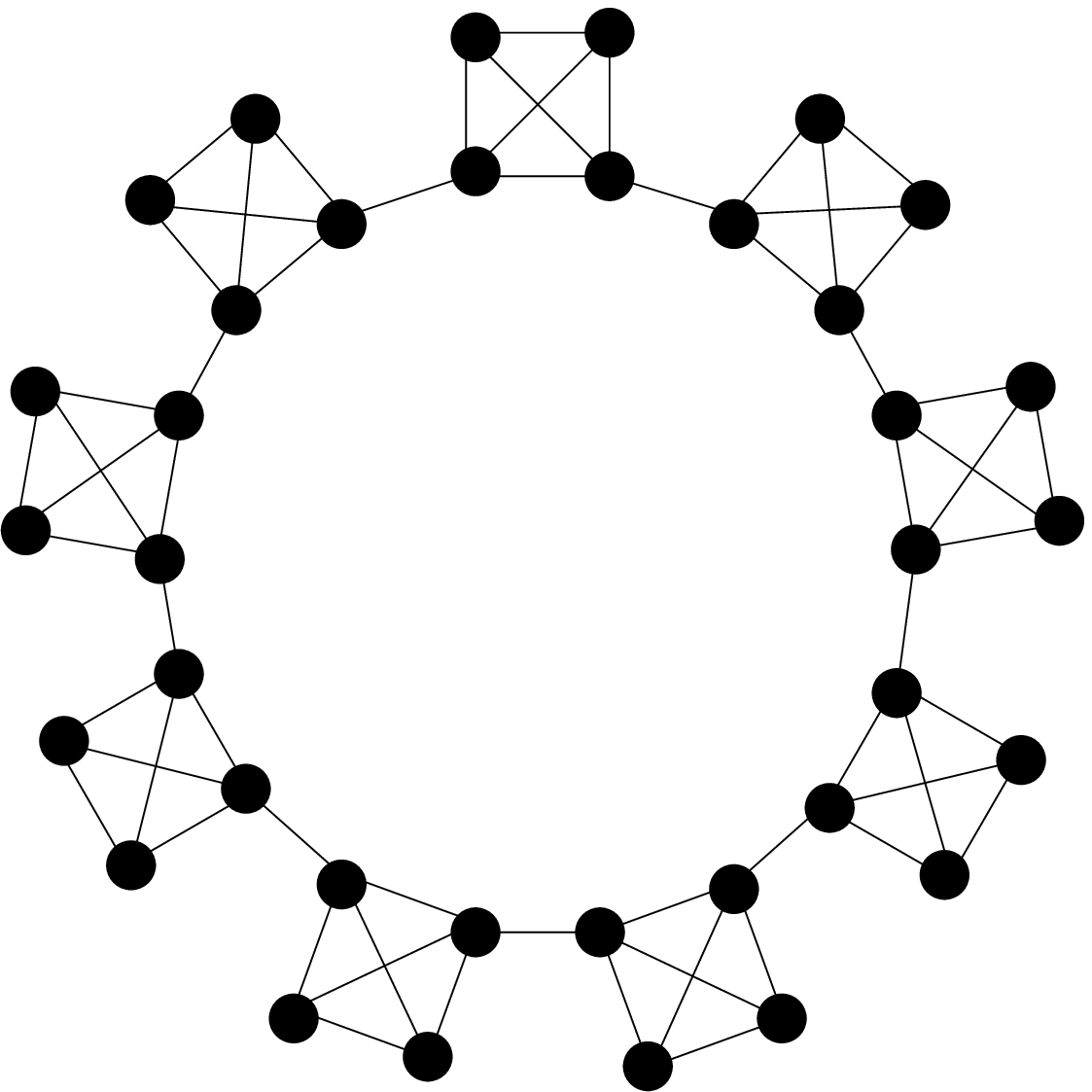} & 
\hspace{2cm}\includegraphics[width=4.0cm]{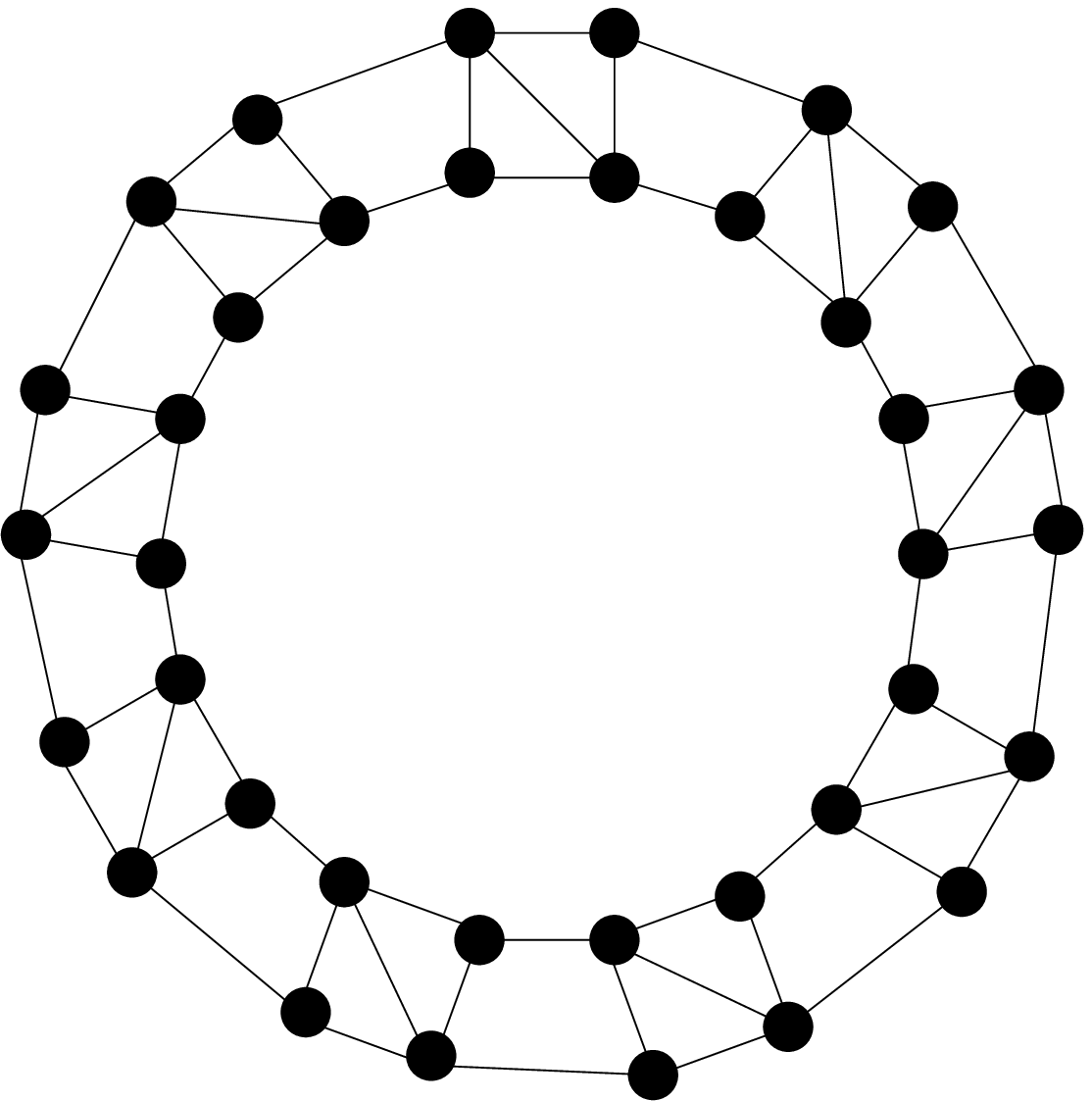} \\
{\LARGE (a)} & \hspace{2cm} {\LARGE (b)} \\ 
 & \\
\includegraphics[width=4.0cm]{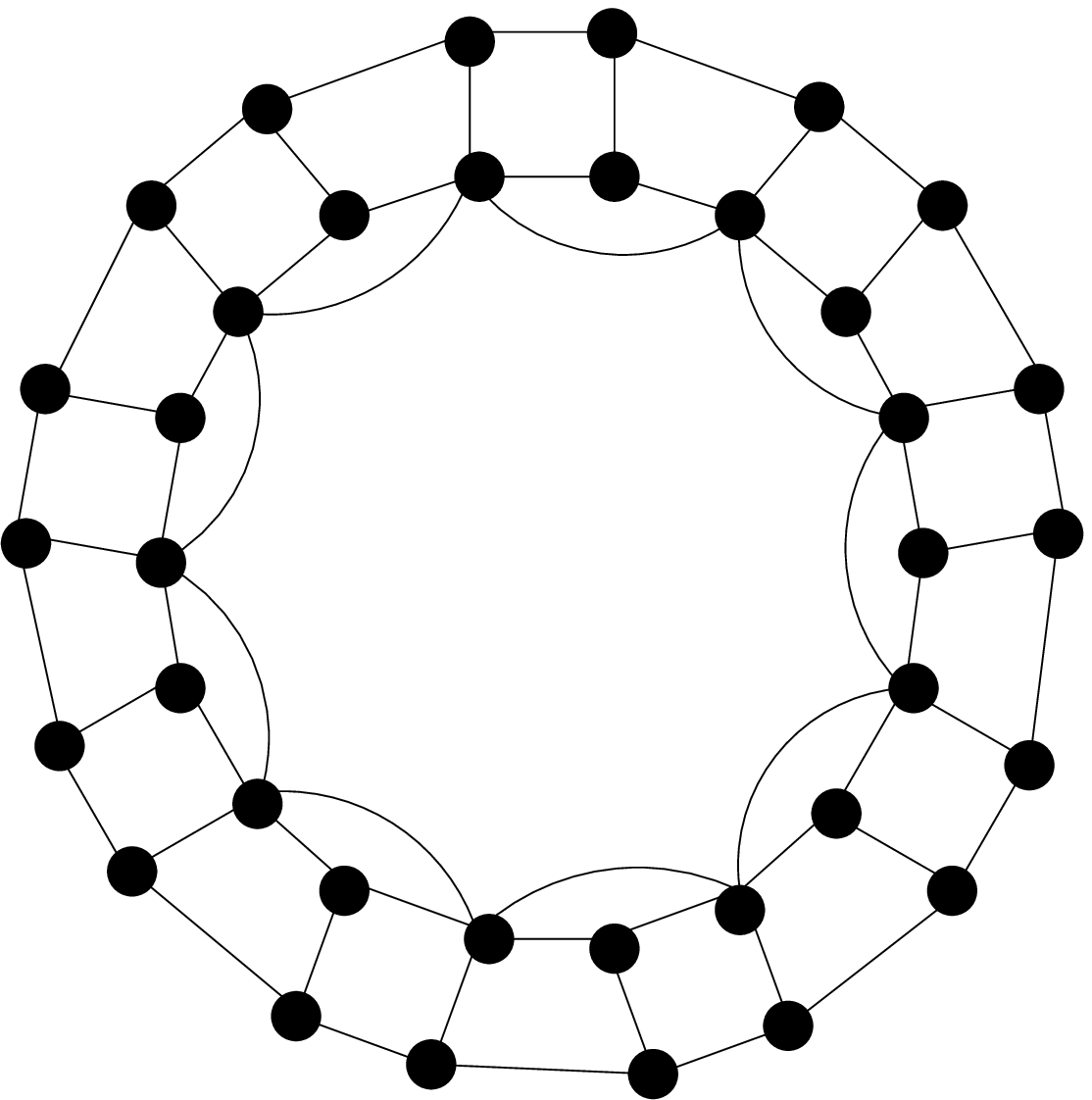} & 
\hspace{2cm} \includegraphics[width=4.0cm]{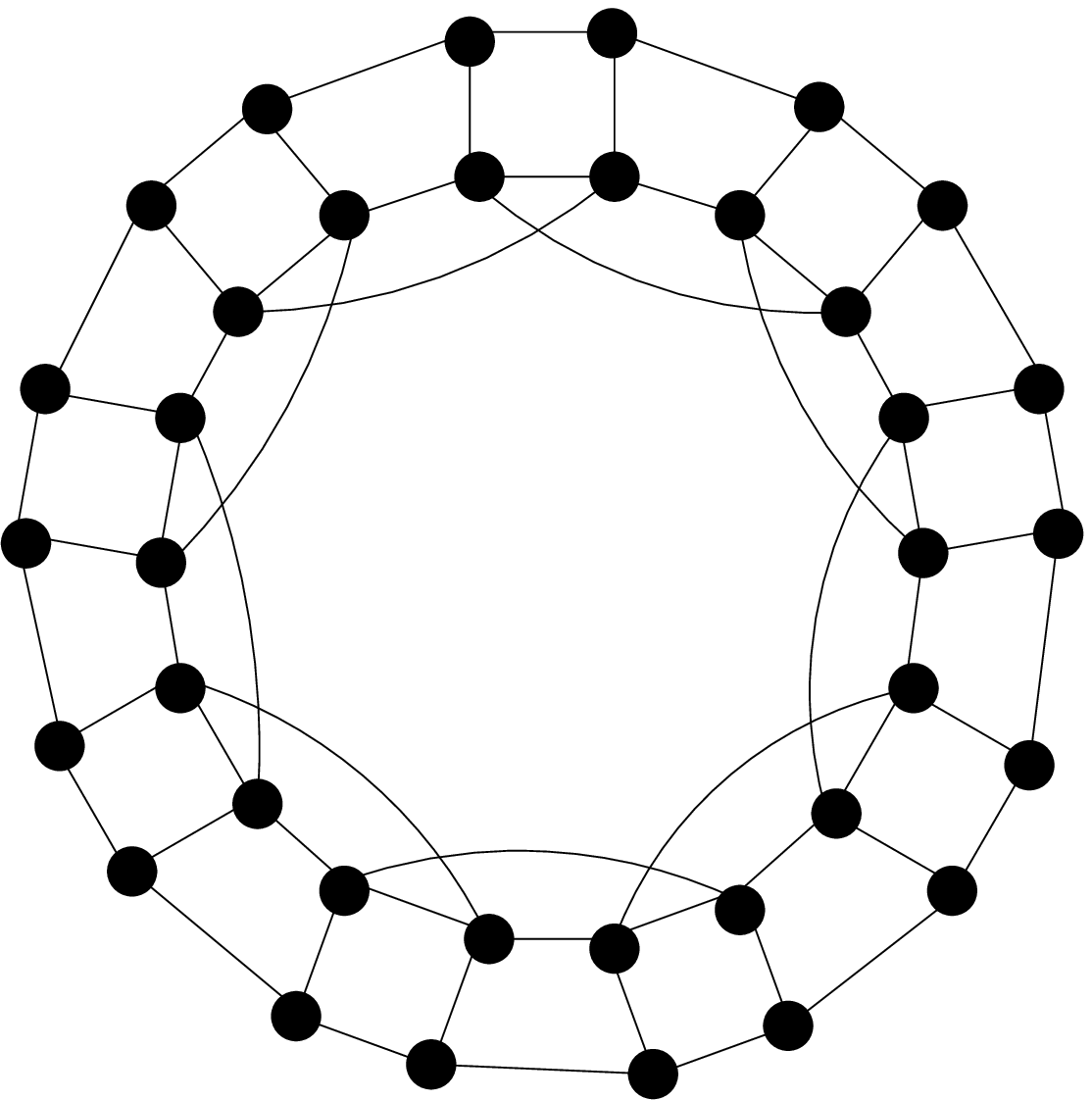} \\
{\LARGE (c)} & \hspace{2cm} {\LARGE (d)} \\
\end{tabular}

\vspace{11.5cm} 
{\Large FIG. 9}
\end{center} 
\end{figure}

\end{document}